\documentclass[sn-mathphys,Numbered]{sn-jnl}

\usepackage{epsfig,amsmath,amssymb,bm}
\usepackage[legacycolonsymbols]{mathtools}
\usepackage{typearea}\typearea{13}
\usepackage{graphicx}%
\usepackage{multirow}%
\usepackage{amsmath,amssymb,amsfonts}%
\usepackage{amsthm}%
\usepackage{mathrsfs}%
\usepackage[title]{appendix}%
\usepackage{xcolor}%
\usepackage{textcomp}%
\usepackage{manyfoot}%
\usepackage{booktabs}%
\usepackage{algorithm}%
\usepackage{algorithmicx}%
\usepackage{algpseudocode}%
\usepackage{listings}%

\newtheorem{prop}{Prop}[section]
\newtheorem{corollary}[prop]{Corollary} 
\newtheorem{lemma}[prop]{Lemma}

\newtheorem{theorem}[prop]{Theorem}

\makeatletter
\@addtoreset{equation}{section}
\makeatother
\renewcommand{\theequation}{\thesection.\arabic{equation}}
\makeatletter
\long\def\@makecaption#1#2{{\small
\advance\leftskip1cm
\advance\rightskip1cm
\vskip\abovecaptionskip
\sbox\@tempboxa{#1: #2}%
\ifdim \wd\@tempboxa >\hsize
 #1: #2\par
\else
\global \@minipagefalse
\hb@xt@\hsize{\hfil\box\@tempboxa\hfil}%
\fi
\vskip\belowcaptionskip}}
\makeatother
\def\eq#1\en{\begin{equation}#1\end{equation}}  
\def\eqa#1\ena{\begin{align}#1\end{align}}
\def\eqg#1\eng{\begin{gather}#1\end{gather}}
\newcommand{\lb}[1]{\label{e:#1}}
\newcommand{\rlb}[1]{\eqref{e:#1}} 
\newcommand{\nl}{\notag\\}

\newcommand{\up}{\uparrow}
\newcommand{\dn}{\downarrow}
\newcommand{\bs}{\backslash}
\newcommand{\sumtwo}[2]%
{\mathop{\sum_{#1}}_{#2}}
\newcommand{\sumthree}[3]%
{\mathop{\mathop{\sum_{#1}}_{#2}}_{#3}}
\newcommand{\sumfour}[4]%
{\mathop{\mathop{\mathop{\sum_{#1}}_{#2}}_{#3}}_{#4}} 

\newcommand{\snorm}[1]{\Vert#1\Vert}

\newcommand{\sbkt}[1]{\langle#1\rangle}
\newcommand{\bbkt}[1]{\bigl\langle#1\bigr\rangle}

\newcommand{\bbE}{\mathbb{E}}

\newcommand{\bbR}{\mathbb{R}}
\newcommand{\bbZ}{\mathbb{Z}}

\newcommand{\bszero}{\boldsymbol{0}}
\newcommand{\bsb}{\boldsymbol{b}}
\newcommand{\bstau}{\boldsymbol{\tau}}
\newcommand{\bss}{\boldsymbol{\sigma}}
\newcommand{\bssa}{\boldsymbol{\sigma}_\mathrm{a}}
\newcommand{\bssb}{\boldsymbol{\sigma}_\mathrm{b}}
\newcommand\calB{\mathcal{B}}
\newcommand\calC{\mathcal{C}}

\newcommand\calJ{\mathcal{J}}
\newcommand\calO{\mathcal{O}}

\newcommand\dcalB{\partial\mathcal{B}}
\newcommand\dcalC{\partial\mathcal{C}}

\newcommand{\La}{\Lambda}
\newcommand{\LaL}{\Lambda_L}
\newcommand{\dLaL}{\partial\Lambda_L}
\newcommand{\Lal}{\Lambda_\ell}
\newcommand{\Laa}{\Lambda_\mathrm{a}}
\newcommand{\Lab}{\Lambda_\mathrm{b}}

\newcommand{\limsupL}{\limsup_{L\up\infty}}
\newcommand{\liminfL}{\liminf_{L\up\infty}}
\newcommand{\limL}{\lim_{L\up\infty}}
\newcommand{\limN}{\lim_{N\up\infty}}
\newcommand{\qbr}{q_{\rm br}}
\newcommand{\qEA}{q_{\rm EA}}
\newcommand{\qjump}{q_{\rm jump}}
\newcommand{\qrsb}{q_{\rm lrsb}}
\newcommand{\qmin}{q_{\rm min}}
\newcommand{\qmax}{q_{\rm max}}
\newcommand{\la}{\lambda}
\newcommand{\betac}{\beta_\mathrm{c}}
\newcommand{\ep}{\varepsilon}
\newcommand{\tZ}{\tilde{Z}}

\begin{document}
\title{ 
Griffiths-type theorems for short-range spin glass models}
\author*[1]{\fnm{Chigak} \sur{Itoi}}\email{itoi.chigaku@nihon-u.ac.jp}
\author[2]{\fnm{Hisamitsu} \sur{Mukaida}}\email{mukaida@saitama-med.ac.jp}
\author[3]{\fnm{Hal} \sur{Tasaki}}\email{hal.tasaki@gakushuin.ac.jp}
\affil*[1]{
\orgdiv{Department of Physics}, \orgname{Nihon University}, 
\orgaddress{\street{Kanda-Surugadai, Chiyoda-ku}, \state{Tokyo}, \postcode{101-8308}, \country{Japan}}
}
\affil[2]{\orgdiv{Department of Liberal Arts}, \orgname{Saitama Medical University}, 
\orgaddress{\street{38 Moro-Hongo}, \city{Iruma-gun},\state{Saitama}, \postcode{350-0495}, \country{Japan}}
}
\affil[3]{
\orgdiv{Department of Physics}, \orgname{Gakushuin University}, 
\orgaddress{{Mejiro, Toshima-ku}, \state{Tokyo}, \postcode{171-8588}, \country{Japan}}
}
\abstract{We establish relations between different characterizations of order in spin glass models.
We first prove that the broadening of the replica overlap distribution indicated by a nonzero standard deviation of the replica overlap $R^{1,2}$ implies the non-differentiability of the two-replica free energy with respect to the replica coupling parameter $\la$.
In $\bbZ_2$ invariant models such as the standard Edwards-Anderson model, the non-differentiability is equivalent to the spin glass order characterized by a nonzero Edwards-Anderson order parameter.
This generalization of Griffiths' theorem is proved for any short-range spin glass models with classical bounded spins.
We also prove that the non-differentiability of the two-replica free energy mentioned above implies replica symmetry breaking in the literal sense, i.e., a spontaneous breakdown of the permutation symmetry in the model with three replicas.
This is a general result that applies to a large class of random spin models, including long-range models such as the Sherrington-Kirkpatrick model and the random energy model.
\par\noindent
{\em There is a 25-minute video that explains the main results of the present work:} 
\\\url
{https://youtu.be/BF3hJiY1xvI}
}
\keywords{spin glass, replica symmetry breaking, Edwards-Anderson model, long-range order, overlap of spin configurations}
\maketitle
\setcounter{footnote}{0}

\tableofcontents

\section{Introduction}
\label{s:intro}
In spite of extensive studies over decades, the phase structure of spin glass models is still a widely open problem \cite{BovierBook,MPV,N,T}.
Although some aspects of models with long-range interactions are understood with mathematical rigor, very little has been proved for short-range models.
In fact, the nature of spin glass order in the most basic Edwards-Anderson (EA) model in three
dimensions is still controversial; there has been a long debate about whether the model exhibits replica symmetry breaking (RSB) as is predicted by the mean-field theory \cite{P0,MPV} or does not as is predicted by the droplet theory \cite{FH,FH2}.
In the present paper, we examine the characterization of spin glass order in (mainly) short-range spin glass models and prove rigorous inequalities between different order parameters.

To motivate the present work, let us briefly review the celebrated theorem by Griffiths \cite{Gff} for the ferromagnetic Ising model on the $d$-dimensional $L\times\cdots\times L$ hypercubic lattice $\LaL$ with the Hamiltonian $H(\bss)=-\sum_{\{x,y\}\in\calB_L}\sigma_x\sigma_y$.
See section~\ref{s:single} for notations.
The model has $\bbZ_2$ symmetry in the sense that $H(\bss)$ is invariant under the global spin flip $\sigma_x\to-\sigma_x$ for all $x\in\LaL$.
It is known that the model is in the ferromagnetic phase at sufficiently low temperatures if $d\ge2$.
The ferromagnetic order is conveniently characterized by the long-range order parameter
\eq
\mu_\mathrm{LRO}\coloneqq\limL\sqrt{\frac{1}{L^{2d}}\sum_{x,y\in\LaL}\sbkt{\sigma_x\sigma_y}_{L,\beta}},
\lb{muLRO}
\en
where $\sbkt{\cdots}_{L,\beta}$ denotes the thermal expectation value at inverse temperature $\beta$
for open or periodic boundary conditions.
One has $\mu_\mathrm{LRO}>0$ if $\sbkt{\sigma_x\sigma_y}_{L,\beta}$ does not decay to zero as $|x-y|$ grows.
Another useful order parameter is the spontaneous magnetization
\eq
\mu_\mathrm{SM}\coloneqq\limL\frac{1}{L^d}\sum_{x\in\LaL}\sbkt{\sigma_x}_{L,\beta;+},
\lb{muSM}
\en
where $\sbkt{\cdots}_{L,\beta;+}$ is the expectation value with the plus boundary condition (see \rlb{RSHamil} below).
Note that the global spin-flip symmetry is explicitly broken by the boundary condition.
The spontaneous magnetization $\mu_\mathrm{SM}$ measures possible spontaneous breakdown of the global spin-flip symmetry.
Griffiths \cite{Gff} proved that the two order parameters are related by the inequality
\eq
\mu_\mathrm{SM}\ge\mu_\mathrm{LRO}.
\lb{muSMLRO}
\en
This means the ferromagnetic order characterized by long-range order inevitably implies the ferromagnetic order characterized by nonzero spontaneous magnetization.
Although the corresponding equality $\mu_\mathrm{SM}=\mu_\mathrm{LRO}$ is now known for the ferromagnetic Ising model, where the complete classification of translation invariant equilibrium states has been accomplished \cite{Lebowitz1977,GallavottiMiracleSole1972,Bodineau2006}, the original proof by Griffiths has universal applicability since it relies only on basic properties of statistical mechanics.
Griffiths' theorem was extended to various spin systems, both classical and quantum.
See \cite{KT,Halbook} and references therein.

The major aim of the present work is to properly extend Griffiths' inequality \rlb{muSMLRO} to short-range spin glass models.
The order parameter that corresponds to $\mu_\mathrm{LRO}$ is the broadening order parameter $\qbr$, defined as \rlb{qbr1} or \rlb{qbr2}, which detects the broadening of the probability distribution of the replica overlap $R^{1,2}$.
This is a standard quantity often discussed in connection with the phenomenon of RSB.
In the standard EA model without a magnetic field, the order parameter corresponding to $\mu_\mathrm{SM}$ is the Edward-Anderson (EA) order parameter $\qEA$ \cite{EA,vEG}.
The EA order parameter $\qEA$ is the most traditional order parameter in the theory of spin glasses. 
It detects a possible spontaneous breaking of the $\bbZ_2$ symmetry with respect to the global spin flip in one of the two replicas.
We prove that these order parameters satisfy $\qEA\ge\qbr$ (Theorem~\ref{t:GrZ2}).
This is a direct extension of the inequality \rlb{muSMLRO} of Griffiths'.

For the EA model with a nonzero (random or non-random) magnetic field, it is well known that $\qEA$ does not play the role of an order parameter.
We characterize the corresponding spin glass order by means of the jump order parameter $\qjump$ introduced by van Enter and Griffiths \cite{vEG}, which quantifies the discontinuity in the derivative of the two-replica free energy with respect to the replica coupling parameter $\la$.
In this case, we also prove, through a rigorous inequality, that a nonzero $\qbr$ implies nonzero $\qjump$ (Theorem~\ref{t:Grgen}).
This is again an extension of Griffiths' theorem but has a slightly different nature since the spin glass order in this model is not related to any spontaneous symmetry breaking.
We expect that the present theorem has relevance to the RSB in short-range spin glass models, which is conjectured by  the mean-field theory to take place in sufficiently high enough dimensions.

We should note that these relations between $\qbr$ and $\qEA$ or $\qjump$ are anticipated from a heuristic argument based on the probability distribution of the replica overlap.
See section~\ref{s:dis}.
As far as we know, however, such relations have not been justified rigorously in spin glass models.\footnote{
We recall that rigorous results for short-range random spin systems are still rare.
This is partly because most known correlation inequalities (see, e.g., \cite{SimonBook,FV}) do not hold.
Our main finding is that the simple correlation inequality considered in \cite{HT} leads us to meaningful inequalities between order parameters.
}

Our third theorem, Theorem~\ref{t:rsb}, also relates different characterizations of spin glass order but has a different nature.
It applies to a wide class of spin glass models, both short-range and long-range models, and shows that the non-differentiability of the two-replica free energy (indicated by nonzero $\qjump$) implies the spontaneous breakdown of the replica permutation symmetry in the three-replica system.
This theorem is most meaningful when applied to spin glass models under a nonzero magnetic field, which lacks the $\bbZ_2$ symmetry.
With Theorem~\ref{t:Grgen}, it shows that the broadening of the distribution of the replica overlap, which is usually regarded as a sign of RSB, inevitably leads to ``literal replica symmetry breaking".

All the theorems in the present paper can be proved for a wide class of spin glass models.
The only essential features are that the spins are classical and bounded, and the interactions are short-range and (stochastically) translation invariant.
However, for notational simplicity, we only discuss the Ising-type EA model with nearest neighbor interactions throughout the present paper (except for Theorem~\ref{t:rsb} and Appendices).

\medskip
The present paper is organized as follows.
After carefully defining the models and basic quantities in section~\ref{s:Def}, we discuss our three theorems in sections~\ref{s:GrZ2}, \ref{s:nonZ2}, and \ref{s:LRSB}.
The theorems, except for Theorem~\ref{t:rsb}, are proved separately in section~\ref{s:proof}.
We give some detailed discussions on the random field Ising model and the random energy model in Appendices~\ref{s:RFIM} and \ref{s:REM}, respectively.
In Appendix~\ref{s:gen1st}, we discuss a Griffiths-type theorem for a general first-order phase transition that can be proved by extending the proof of Theorem~\ref{t:Grgen}.

\section{Definitions}
 \label{s:Def}
\subsection{Single system}
\label{s:single}
\begin{figure}
\begin{center}
{\includegraphics[width=5truecm]{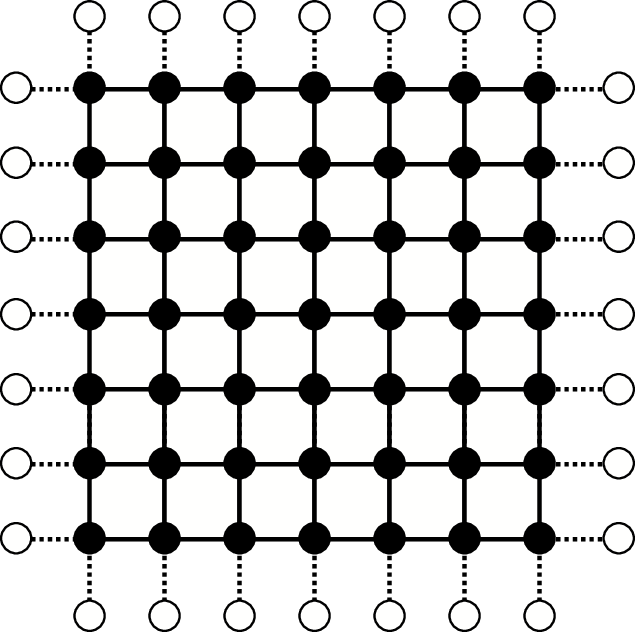}}
\caption[dummy]{
Black dots and white dots represent the sites in $\La_7$ and its boundary $\partial\La_7$, respectively.
Solid lines and dashed lines represent bonds in $\calB_7$ and $\partial\calB_7$, respectively.
}
\label{f:La7}
\end{center}
\end{figure}

For $d=1,2,\ldots$, we regard $\bbZ^d$ as the infinite $d$-dimensional hypercubic lattice, and denote its elements, i.e. sites, as $x,y\ldots\in\bbZ^d$.
The distance between two sites $x,y\in\bbZ^d$ is defined as
\eq
|x-y|\coloneqq\snorm{x-y}_1=\sum_{i=1}^d|x_i-y_i|,
\en
where we wrote $x=(x_1,\ldots,x_d)$.
For a positive integer  $L$ , we consider the $d$-dimensional $L\times\ldots L$ hypercubic lattice
\eq
\LaL\coloneqq\{1,\ldots,L\}^d\subset\bbZ^d,
\lb{LaL}
\en
and its boundary
\eq
\dLaL\coloneqq\bigl\{u\in\bbZ^d\bs\LaL\,\bigl|\,|u-x|=1\ \text{for some}\ x\in\LaL\bigr\}.
\en
The set of bonds, i.e., unordered sets of neighboring sites in $\LaL$ is denoted as
\eq
\calB_L\coloneqq\bigl\{\{x,y\}\,|\,x,y\in\LaL,\ |x-y|=1\bigr\},
\en
and the set of boundary bonds, i.e., oriented pairs of neighboring sites in $\LaL$ and $\dLaL$ as
\eq
\dcalB_L\coloneqq\bigl\{(x,u)\,\bigl|\, x\in\LaL,\ u\in\dLaL,\ |x-u|=1\,\bigr\}.
\en
See Figure~\ref{f:La7}.

We associate each site $x\in\LaL$ with an Ising spin described by the spin variable $\sigma_x\in\{1,-1\}$.
A spin configuration, i.e., the collection of all spin variables on $\LaL$, is denoted as $\bss=(\sigma_x)_{x\in\LaL}$.
By $\calC_L$, we denote the set of all spin configurations on $\LaL$.
Similarly, we associate each boundary site $u\in\dLaL$ with the boundary ``spin'' described by a continuous variable $b_u\in[-1,1]$.
The collection of all $b_u$ for $u\in\dLaL$ is denoted as $\bsb=(b_u)_{u\in\dLaL}$, and the set of all $\bsb$ is denoted as $\dcalC_L$.

For any $x,y\in\bbZ^d$ such that $|x-y|=1$, let $J_{x,y}=J_{y,x}\in\bbR$ be the random exchange interaction between sites $x$ and $y$, and for any $x\in\bbZ^d$, let $h_x\in\bbR$ be the random magnetic field at site $x$.
We assume that all $J_{x,y}$ are independent and identically distributed and that all $h_x$ are independent and identically distributed.
The probability distributions are arbitrary, except for the assumptions
\eq
\bbE\,|J_{x,y}|<\infty,\quad\bbE\,|h_x|<\infty,
\lb{EJ}
\en
where $\bbE\,f$ denotes the expectation value of $f$.
Throughout the present paper, we assume that the distributions for $J_{x,y}$ and $h_x$ are fixed.
We thus do not make explicit the dependence of various quantities on the distributions.

As limiting cases, we can assume either the interaction or the magnetic field is non-random.
If the interaction is non-random, i.e., $J_{x,y}=J$ for any $x,y\in\bbZ^d$ such that $|x-y|=1$, the model reduces to the random field Ising model.
If the magnetic field is non-random, i.e., $h_x=h$ for all $x\in\bbZ^d$, we have the EA model under a uniform magnetic field.\footnote{%
By a slight modification of the proof, we can also treat an interaction and a magnetic field that are periodic.}

Let us denote by $\calJ_L$ the collection of all $J_{x,y}$ such that $x\in\LaL$ or $y\in\LaL$ and all $h_x$ such that   $x\in\LaL$.
Then the Hamiltonian of the Edwards-Anderson (EA) model \cite{EA} with a random magnetic field is defined by
\begin{equation}
H_L(\bss;\calJ_L,\bsb)\coloneqq - \sum_{\{x,y\}\in\calB_L} J_{x,y}\,\sigma_x \sigma_y
-\sum_{(x,u)\in \dcalB_L}J_{x,u}\,\sigma_x b_u
- \sum_{x \in \LaL}h_x\,\sigma_x,
\lb{RSHamil}
\end{equation}
where $\bsb\in\dcalC_L$ defines the boundary condition of the spin system.
Standard choices are the open boundary condition with $b_u=0$ for all $u\in\dLaL$ and the plus boundary condition with  $b_u=1$ for all $u\in\dLaL$.

We are ready to define thermal expectation values and thermodynamic functions.
Let $F(\bss)$ be an arbitrary function of spin configuration $\bss\in\calC_L$.
Then its thermal expectation value at  inverse temperature $\beta >0$ is defined as
\eq
\sbkt{F}_{L,\beta;\calJ_L,\bsb}
= \frac{1}{Z_L(\beta;\calJ_L,\bsb)} \sum_{\bss \in\calC_L} F(\bss)\,e^{ - \beta H_L(\bss;\calJ_L,\bsb)},
\lb{<F>}
\en
where the partition function is defined by
\begin{equation}
Z_L(\beta;\calJ_L,\bsb) \coloneqq \sum_{\bss \in\calC_L} e^{ - \beta H_L(\bss;\calJ_L,\bsb)}.
\end{equation}
We shall sometimes allow the boundary configuration $\bsb$ to depend on the inverse temperature $\beta$ and the interactions and magnetic fields, $\calJ_L$, and express the dependence explicitly as $\bsb(\beta,\calJ_L)$.
We then define the corresponding averaged free energy (or, to be more precise, the free energy density)
\begin{equation}
f_L(\beta;\bsb(\cdot))\coloneqq- \frac{1}{\beta L^d}\,\bbE\, \log Z_L(\beta;\calJ_L,\bsb(\beta,\calJ_L)), 
\lb{fL}
\end{equation}
and its infinite-volume limit
\eq
f(\beta)\coloneqq\lim_{L\up\infty}f_L(\beta;\bsb(\cdot)).
\lb{flim}
\en
The limit exists and is independent of the boundary condition.
See Lemma~\ref{L:f} below.

\subsection{Replicated systems}
\label{s:rep}
In order to study order in spin glass models, we introduce $n$ replicated spin systems.
For our purpose, it suffices to consider the cases with $n=2$ and 3, which are described by the Hamiltonians
\eq
H^{(2)}_L(\bss^{1},\bss^{2};\la,\calJ_L,\bsb)\coloneqq\sum_{\alpha=1}^2H_L(\bss^{\alpha};\calJ_L,\bsb)-\la\sum_{x\in\LaL}\sigma^{1}_x\sigma^{2}_x,
\lb{H2}
\en
and
\eq
H^{(3)}_L(\bss^{1},\bss^{2},\bss^{3};\la,\la',\calJ_L,\bsb)\coloneqq\sum_{\alpha=1}^3H_L(\bss^{\alpha};\calJ_L,\bsb)-\la\sum_{x\in\LaL}\sigma^{1}_x\sigma^{2}_x-\la'\sum_{x\in\LaL}\sigma^{1}_x\sigma^{3}_x,
\lb{H3}
\en
respectively, where $\bss^{\nu}=(\sigma_x^{\nu})_{x\in\LaL}\in\calC_L$ denotes a spin configuration in the $\nu$-th replica.
Here we did not simply replicate the same system but also (artificially) introduced explicit couplings between different replicas with coupling parameters $\la,\la'\in\bbR$ in order to test for possible spin glass order.
We then define the expectation values as
\eqg
\sbkt{F}^{(2)}_{L,\beta,\la;\calJ_L,\bsb}
= \frac{1}{Z^{(2)}_L(\beta,\la;\calJ_L,\bsb)} \sum_{\bss^{1},\bss^{2}\in\calC_L} F(\bss^{1},\bss^{2})\,e^{ - \beta H^{(2)}_L(\bss^{1},\bss^{2};\la,\calJ_L,\bsb)},
\lb{F2}\\
\sbkt{F}^{(3)}_{L,\beta,\la,\la';\calJ_L,\bsb}
= \frac{1}{Z^{(3)}_L(\beta,\la,\la';\calJ_L,\bsb)} \sum_{\bss^{1},\bss^{2},\bss^{3}\in\calC_L} F(\bss^{1},\bss^{2},\bss^{3})\,e^{ - \beta H^{(3)}_L(\bss^{1},\bss^{2},\bss^{3};\la,\la',\calJ_L,\bsb)},
\eng
with the partition functions
\eqg
Z^{(2)}_L(\beta,\la;\calJ_L,\bsb)=\sum_{\bss^{1},\bss^{2}\in\calC_L} 
e^{ - \beta H^{(2)}_L(\bss^{1},\bss^{2};\la,\calJ_L,\bsb)},
\lb{ZL2}\\
Z^{(3)}_L(\beta,\la,\la';\calJ_L,\bsb)= \sum_{\bss^{1},\bss^{2},\bss^{3}\in\calC_L} 
e^{ - \beta H^{(3)}_L(\bss^{1},\bss^{2},\bss^{3};\la,\la',\calJ_L,\bsb)}.
\lb{ZL3}
\eng
We again define the averaged free energy for the replicated systems as
\eqg
f^{(2)}_L(\beta,\la;\bsb(\cdot))\coloneqq- \frac{1}{\beta L^d}\,\bbE\, \log Z^{(2)}_L(\beta,\la;\calJ_L,\bsb(\beta,\calJ_L)), 
\lb{fL2}\\
f^{(3)}_L(\beta,\la,\la';\bsb(\cdot))\coloneqq- \frac{1}{\beta L^d}\,\bbE\, \log Z^{(3)}_L(\beta,\la,\la';\calJ_L,\bsb(\beta,\calJ_L)).
\lb{fL3}
\eng
Note that the permutation symmetry between replicas imply
\eq
f^{(3)}_L(\beta,\la,\la';\bsb(\cdot))=f^{(3)}_L(\beta,\la',\la;\bsb(\cdot)).
\lb{fp}
\en
We also see that
\eqg
f^{(2)}_L(\beta,0;\bsb(\cdot))=2f_L(\beta;\bsb(\cdot)),
\lb{f2}\\
f^{(3)}_L(\beta,\la,0;\bsb(\cdot))=f_L(\beta;\bsb(\cdot))+f^{(2)}_L(\beta,\la;\bsb(\cdot)).
\lb{f3}
\eng
Straightforward calculations verify that
\eqg
\frac{\partial}{\partial\la}f^{(2)}_L(\beta,\la;\bsb(\cdot))=-\bbE\,\sbkt{R_L^{1,2}}^{(2)}_{L,\beta,\la;\calJ_L,\bsb(\beta,\calJ_L)},
\lb{f2la}\\
\frac{\partial^2}{\partial\la^2}f^{(2)}_L(\beta,\la;\bsb(\cdot))=-\beta L^d\,\bbE\Bigl\{\bbkt{(R_L^{1,2})^2}^{(2)}_{L,\beta,\la;\calJ_L,\bsb(\beta,\calJ_L)}-\bigl(\sbkt{R_L^{1,2}}^{(2)}_{L,\beta,\la;\calJ_L,\bsb(\beta,\calJ_L)}\bigr)^2\Bigr\}\le0,
\lb{f2la2}
\eng
where we defined the overlap of two spin configurations by
\eq
R_L^{\alpha,\beta}(\bss^{\alpha},\bss^{\beta})\coloneqq\frac{1}{L^d}\sum_{x\in\LaL}\sigma^\alpha_x\sigma^\beta_x,
\lb{Rab}
\en
for $\alpha\ne\beta$.
From \rlb{f2la2} we see that $f^{(2)}_L(\beta,\la;\bsb(\cdot))$ is concave (or convex-upward) in $\la\in\bbR$.
We can similarly show that
\eq
\Bigl(\xi\frac{\partial}{\partial\la}+\xi'\frac{\partial}{\partial\la'}\Bigr)f^{(3)}_L(\beta,\la,\la';\bsb(\cdot))
=-\bbE\,\sbkt{\tilde{R}}^{(3)}_{L,\beta,\la,\la';\calJ_L,\bsb(\beta,\calJ_L)},
\lb{f3la}
\en
\eqa
\Bigl(\xi\frac{\partial}{\partial\la}+\xi'\frac{\partial}{\partial\la'}\Bigr)^2&f^{(3)}_L(\beta,\la,\la';\bsb(\cdot))
\nl&=-\beta L^d\,\bbE\Bigl\{\bbkt{(\tilde{R})^2}^{(3)}_{L,\beta,\la,\la';\calJ_L,\bsb(\beta,\calJ_L)}-\bigl(\sbkt{\tilde{R}}^{(3)}_{L,\beta,\la,\la';\calJ_L,\bsb(\beta,\calJ_L)}\bigr)^2\Bigr\}\le0,
\lb{f3la2}
\ena
with $\tilde{R}=\xi\,R_L^{1,2}+\xi'\,R_L^{1,3}$ for any $\xi,\xi'\in\bbR$.
Again \rlb{f3la2} implies that $f^{(3)}_L(\beta,\la,\la';\bsb(\cdot))$ is concave in $(\la,\la')\in\bbR^2$.

Let us define the infinite-volume limits of the free energies
\eqg
f^{(2)}(\beta,\la)=\lim_{L\up\infty}f^{(2)}_L(\beta,\la;\bsb(\cdot)),\lb{f2lim}\\
f^{(3)}(\beta,\la,\la')=\lim_{L\up\infty}f^{(3)}_L(\beta,\la,\la';\bsb(\cdot)),\lb{f3lim}
\eng
whose existence and independence on the boundary conditions are guaranteed as follows.
\begin{lemma}[Infinite-volume limits of the free energy densities]\label{L:f}
The limits \rlb{flim}, \rlb{f2lim}, and \rlb{f3lim} exist and are independent of the boundary condition $\bsb(\cdot)$.
\end{lemma}
The lemma can be proved by employing the standard technique.
See, e.g., \cite{SimonBook,BovierBook,FV}.
It is crucial for us that $f^{(2)}(\beta,\la)$ is concave in $\la\in\bbR$ and $f^{(3)}(\beta,\la,\la')$ is concave in $(\la,\la')\in\bbR^2$. 

Finally, it may be convenient to define the distribution of the replica overlap, although we do not discuss this quantity in detail.
Define the characteristic function by $\chi[\text{true}]=1$ and $\chi[\text{false}]=0$.
Then $P_{L,\beta}(q)$ for $q\in[-1,1]$ is defined by
\eq
\int_{-1}^q dr\,P_{L,\beta}(r)=\bbE\bbkt{\,\chi[R^{1,2}(\bss^1,\bss^2)\le q]\,}_{L,\beta,0;\calJ_L,\bszero}^{(2)}.
\lb{PLq}
\en
Clearly, the limit $P_\beta(q)=\limL P_{L,\beta}(q)$ (if exists) describes the probability density that the overlap $R^{1,2}$ is identical to $q$ in the infinite-volume ensemble.

\section{Characterizations of spin glass order}
\label{s:main}
Let us discuss our main results on characterizations of order in spin glass models.
We note that all the results are most meaningful when our spin systems do not exhibit standard order, such as the ferromagnetic or the antiferromagnetic order.
Our aim is to clarify relations between different notions of spin glass order by means of rigorous and general inequalities between different order parameters.
The central objects are the replica overlap $R_L^{\alpha,\beta}(\bss^{\alpha},\bss^{\beta})$ defined in \rlb{Rab}, and the two-replica and the three-replica free energies defined by \rlb{ZL2}, \rlb{ZL3}, \rlb{fL2}, \rlb{fL3}, \rlb{f2lim}, and \rlb{f3lim}.

\subsection{Griffiths' theorem for spin glass models with $\bbZ_2$ symmetry}
\label{s:GrZ2}
We shall first focus on the EA model \rlb{RSHamil} without a magnetic field, i.e., $h_x=0$ for all $x\in\bbZ^d$.
This is the original model proposed by Edwards and Anderson \cite{EA}.
It is believed that the model exhibits spin glass order at sufficiently low temperatures if $d\ge3$. 
However, the nature of the spin glass phase, especially the presence or absence of RSB, is still controversial.
See, e.g., \cite{MPV,FH,FH2,NS}.
It is essential to note that the model has global $\bbZ_2$ symmetry in the sense that the Hamiltonian (with the open boundary condition) is invariant under the global spin flip, $\sigma_x\to-\sigma_x$ for all $x$.

Let us discuss the characterization of spin glass order using the system with two replicas whose Hamiltonian is \rlb{H2}.
We first treat the model with the open boundary condition\footnote{%
One can treat more general boundary conditions, including the periodic boundary condition by a straightforward extension.
See setion~\ref{s:BC}.
} realized by $\bsb=\bszero\coloneqq(0,\ldots,0)$ and the vanishing replica coupling parameter $\la=0$.
In this case, the model is invariant under the global spin-flip applied to each replica and thus has the $\bbZ_2\times\bbZ_2$ symmetry.
We are particularly interested in the spin-flip applied to one of the two replicas, i.e.,
\eq
\sigma_x^{(1)}\to\sigma_x^{(1)},\ \sigma_x^{(2)}\to-\sigma_x^{(2)}\quad\text{for all $x\in\LaL$}.
\lb{Z2}
\en
Under the transformation \rlb{Z2}, the Hamiltonian $H^{(2)}_L(\bss^{1},\bss^{2};0,\calJ_L,\bszero)$ is invariant while the replica overlap $R_L^{1,2}(\bss^{1},\bss^{2})$ changes its sign.
We thus find
\eq
\sbkt{R_L^{1,2}}^{(2)}_{L,\beta,0;\calJ_L,\bszero}=0,
\lb{R12=0}
\en
for any $\beta$ and $\calJ_L$.
This suggests that spin glass order should be characterized by the fluctuation of $R_L^{1,2}$ or, equivalently, the broadening of the replica overlap distribution $P_\beta(q)$ (see \rlb{PLq}).
We thus define the order parameter for broadening as
\eq
\qbr\coloneqq\limsupL\sqrt{\bbE\,\bbkt{(R_L^{1,2})^2}^{(2)}_{L,\beta,0;\calJ_L,\bszero}}
=\limsupL\sqrt{\frac{1}{L^{2d}}\,\bbE\sum_{x,y\in\LaL}\bigl(\sbkt{\sigma_x\sigma_y}_{L,\beta;\calJ_L,\bszero}\bigr)^2}.
\lb{qbr1}
\en

A more common order parameter for spin glass is the Edwards-Anderson (EA) order parameter, which detects a possible spontaneous breakdown of the $\bbZ_2$ symmetry \rlb{Z2}.
It is usually written as
\eq
\qEA\overset{?}{=}\bbE\,\sbkt{\sigma_x}^2=\bbE\,\sbkt{R^{1,2}}^{(2)},
\en
but this expression is ambiguous about what the thermal expectation values $\sbkt{\sigma_x}$ or $\sbkt{R^{1,2}}^{(2)}$ precisely mean.
From the poor notation, one might erroneously conclude that $\qEA$ is always vanishing because of \rlb{R12=0}.
In order to define the EA order parameter that properly characterizes spin glass order, one must break the $\bbZ_2$ symmetry \rlb{Z2} explicitly but infinitesimally.

A physically natural definition of the EA order parameter was given by van Enter and Griffits \cite{vEG}.
For any $L$, let
\eq
\qEA(L)\coloneqq\bbE\max_{\bsb\in\dcalC_L}\sbkt{R_L^{1,2}}^{(2)}_{L,\beta,0;\calJ_L,\bsb}
=\frac{1}{L^d}\bbE\max_{\bsb\in\dcalC_L}\sum_{x\in\LaL}\bigl(\sbkt{\sigma_x}_{L,\beta;\calJ_L,\bsb}\bigr)^2,
\lb{qL}
\en
where we choose a boundary configuration $\bsb$ that maximizes the overlap $\sbkt{R_L^{1,2}}^{(2)}_{L,\beta,0;\calJ_L,\bsb}$ for each combination of $\beta$, $L$, and $\calJ_L$.
This means $\bsb$ generally depends on $L$, $\beta$, and $\calJ_L$.
Note that the $\bbZ_2$ symmetry \rlb{Z2} is explicitly broken by the boundary condition.
Then the EA order parameter is defined as the infinite-volume limit 
\eq
\qEA\coloneqq\limL\qEA(L).
\lb{qEA1}
\en
The existence of the limit is proved at the end of section~\ref{s:proofmain}.
van Enter and Griffiths \cite{vEG} proved that the same order parameter is expressed as
\eq
\qEA=-\lim_{\la\dn0}\frac{\partial}{\partial\la}f^{(2)}(\beta,\la),
\lb{qEAf}
\en
where  $f^{(2)}(\beta,\la)$ is the two-replica free energy with explicit replica coupling parameter $\la$ defined in  \rlb{ZL2}, \rlb{fL2}, and \rlb{f2}.
Recall that concavity implies $f^{(2)}(\beta,\la)$ is differentiable in $\la$ except at a countable number of points, and hence the limit in the right-hand side of \rlb{qEAf} exists although the derivative $\partial f^{(2)}(\beta,\la)/\partial\la$ may not exist for all $\la$; the limit coincides with the right derivative of $f^{(2)}(\beta,\la)$ at $\la=0$.
We also see from concavity that, for any boundary condition $\bsb(\cdot)$, the derivative $\frac{\partial}{\partial\la}f^{(2)}_L(\beta,\la;\bsb(\cdot))$ converges to $\frac{\partial}{\partial\la}f^{(2)}(\beta,\la)$ as $L\up\infty$ whenever the latter derivative exists.
Then \rlb{f2la} implies the identity
\eq
\qEA=\lim_{\la\dn0}\limL\bbE\sbkt{R_L^{1,2}}^{(2)}_{L,\beta,\la;\calJ_L,\bsb(\beta,\calJ_L)},
\lb{qEAEE}
\en
for any boundary configuration $\bsb(\cdot)$. 
In the expressions \rlb{qEAf} and \rlb{qEAEE} of the EA order parameter, the $\bbZ_2$ symmetry \rlb{Z2} is explicitly broken by the coupling $\la>0$, which is brought to zero after the infinite-volume limit.
Finally, recalling that the global $\bbZ_2$ symmetry \rlb{Z2} implies
\eq
f^{(2)}(\beta,\la)=f^{(2)}(\beta,-\la),
\lb{fsym}
\en
we see from \rlb{qEAf} that
\eq
\qEA=\frac{1}{2}\Bigl\{-\lim_{\la\dn0}\frac{\partial}{\partial\la}f^{(2)}(\beta,\la)+\lim_{\la\up0}\frac{\partial}{\partial\la}f^{(2)}(\beta,\la)\Bigr\}.
\lb{qEAjump}
\en
Thus the EA order parameter also represents the jump in the derivative of $f^{(2)}(\beta,\la)$ at $\la=0$.

Our main result for the $\bbZ_2$ invariant model is the following.
\begin{theorem}[Griffiths' theorem for $\bbZ_2$ invariant model]\label{t:GrZ2}
It holds that
\eq
\qEA\ge\qbr.
\lb{GrZ2}
\en
\end{theorem}

We also note that the symmetry \rlb{fsym}, with \rlb{qEAf}, \rlb{qEAEE}, and \rlb{GrZ2}, implies
\eq
-\lim_{\la\up0}\frac{\partial}{\partial\la}f^{(2)}(\beta,\la)
=\lim_{\la\up0}\limL\bbE\sbkt{R_L^{1,2}}^{(2)}_{L,\beta,\la;\calJ_L,\bsb(\beta,\calJ_L)}
\le-\qbr,
\lb{GrZ2D}
\en
for any boundary configuration $\bsb(\cdot)$. 

From the inequality \rlb{GrZ2}, we see that the assumption $\qbr>0$ leads to the bound $\qEA>0$.
We have thus established that the spin glass order characterized by the broadening of the overlap distribution inevitably implies the spin glass order indicated by a nonzero EA order parameter.
The latter is equivalent to a spontaneous breakdown of the $\bbZ_2$ symmetry \rlb{Z2}.
Note that this is a straightforward generalization of Griffiths' theorem that we discussed in section~\ref{s:intro}.

We note that the present theorem does not provide any information about the origin of the $\bbZ_2$ symmetry breaking.
In particular, one has $\qbr>0$ and hence $\qEA>0$ when the global spin-flip of the single system (rather than the replicated system) is spontaneously broken.
A trivial example is the ferromagnetic Ising model without a magnetic field at low temperatures.

Like the original inequality \rlb{muSMLRO} by Griffiths, our inequality $\qEA\ge\qbr$ is optimal.\footnote{%
The corresponding equality should hold in the ferromagnetic Ising model.}
In fact, a heuristic argument based on the probability distribution of the replica overlap suggests that the corresponding equality $\qEA=\qbr$ holds if and only if the distribution $P_\beta(q)$ has two symmetric peaks at $\pm\qEA$.
See section~\ref{s:dis}.
Recall that the presence of two symmetric peaks is predicted by the droplet theory \cite{FH,FH2} but not by the RSB picture \cite{P0,MPV}.

\subsection{Griffiths-type theorem for spin glass models without $\bbZ_2$ symmetry}
\label{s:nonZ2}
Let us now turn to the general EA model \rlb{RSHamil} with a nonzero magnetic field (which may be random or non-random).
The model lacks the global $\bbZ_2$ symmetry.
In such a model, it is easily observed that the order parameters $\qbr$ and $\qEA$ defined in \rlb{qbr1} and \rlb{qEAf}, respectively, are nonzero even in a totally disordered phase realized at high temperatures.
Roughly speaking, this is because the local magnetic field $h_x$ determines a preferred direction for each spin and leads to a positive correlation between $\sigma_x^{1}$ and $\sigma_x^{2}$.
It is nevertheless possible that a spin glass model without global $\bbZ_2$ symmetry exhibits a spin glass phase, as has been demonstrated in the Sherrington-Kirkpatrick (SK) model with a nonzero random magnetic field \cite{Pn,T}.

Whether a short-range spin glass model without $\bbZ_2$ symmetry has a similar spin glass phase is still a controversial problem. 
The prediction based on the mean-field theory \cite{MPV} is that such a model also possesses a spin glass phase analogous to that in the region bounded by the Almeida-Thouless (AT) line \cite{AT} in the long-range model \cite{P0}.
The droplet theory, on the other hand, predicts that there is no spin glass phase \cite{FH,FH2}. 
Numerical results in three and four dimensions support this picture \cite{BJ,SHYT}.
Then, a plausible scenario may be that there is no spin glass phase in low dimensions, including $d=3$, while there may be spin glass phase in higher dimensions, probably $d>6$.

In models without $\bbZ_2$ symmetry, spin glass order is expected to be characterized by the broadening of the distribution $P_\beta(q)$ of the replica overlap $R_L^{1,2}$ defined in \rlb{PLq}.
The broadening is usually regarded as a sign of RSB.
Since $R_L^{1,2}$ has a nonzero expectation value, the broadening should be characterized by the standard deviation
\eq
\qbr\coloneqq\limsupL\sqrt{
\bbE\,\bbkt{(R_L^{1,2})^2}^{(2)}_{L,\beta,0;\calJ_L,\bszero}
-\Bigl(\bbE\,\bbkt{R_L^{1,2}}^{(2)}_{L,\beta,0;\calJ_L,\bszero}\Bigr)^2
}.
\lb{qbr2}
\en
For a $\bbZ_2$ symmetric model, this coincides with the previous definition \rlb{qbr1} because $\bbE\,\bbkt{R_L^{1,2}}^{(2)}_{L,\beta,0;\calJ_L,\bszero}=0$.
We recall that \rlb{qbr2} is the quantity studied by Chatterjee in \cite{Chatterjee2015} to demonstrate the absence of RSB in the random field Ising model.

We note that the open boundary condition employed in \rlb{qbr2} may be replaced by much more general boundary conditions.
See section~\ref{s:BC}.
We also emphasize that the broadening generally depends on the choice of boundary conditions.
In fact, it can be shown that the quantity corresponding to $\qbr$ is always vanishing in a specific boundary condition.
See the remark at the end of section~\ref{s:proofmain}.

To define the counterpart of the EA order parameter, let us recall the relation \rlb{qEAjump} for a $\bbZ_2$ symmetric model, which shows the EA order parameter can be interpreted as the discontinuity, or jump, in the left and right derivatives of the two-replica free energy.
Following van Enter and Griffiths \cite{vEG}, we introduce the order parameter for models without $\bbZ_2$ symmetry that measures the jump in the derivative as
\eq
\qjump\coloneqq\frac{1}{2}\Bigl\{-\lim_{\la\dn0}\frac{\partial}{\partial\la}f^{(2)}(\beta,\la)+\lim_{\la\up0}\frac{\partial}{\partial\la}f^{(2)}(\beta,\la)\Bigr\}.
\lb{qjump}
\en
Thus nonzero $\qjump$ implies non-differentiability of the two-replica free energy $f^{(2)}(\beta,\la)$ at $\la=0$.
Exactly as in \rlb{qEAEE}, we can express $\qjump$ as
\eq
\qjump=\frac{1}{2}\Bigl\{\lim_{\la\dn0}\limL\bbE\sbkt{R_L^{1,2}}^{(2)}_{L,\beta,\la;\calJ_L,\bsb(\beta,\calJ_L)}
-\lim_{\la\up0}\limL\bbE\sbkt{R_L^{1,2}}^{(2)}_{L,\beta,\la;\calJ_L,\bsb'(\beta,\calJ_L)}\Bigl\},
\lb{GrE}
\en
for any boundary conditions $\bsb(\cdot)$ and $\bsb'(\cdot)$.
One thus sees that $\qjump$ is literally the jump in the expectation value of the overlap $R^{1,2}$ when $\la$ changes from $-0$ to $+0$.
In a $\bbZ_2$ symmetric model, $\qjump$, of course, reduces to the EA order parameter $\qEA$.
It is also possible to express $\qjump$ as in \rlb{qL} and \rlb{qEA1} by using the expectation values with suitably chosen boundary conditions.
See section~V of \cite{vEG}.
See also \rlb{qjEG} below.

We are ready to state our main theorem.
\begin{theorem}[Griffiths-type theorem for the general EA model]\label{t:Grgen}
It holds that
\eq
\qjump\ge\frac{(\qbr)^2}{4}.
\lb{Grqjqbr}
\en
\end{theorem}

The theorem states that the assumption $\qbr>0$ leads to the bound $\qjump>0$.
We have thus established that the spin glass order (or RSB) characterized by the broadening of the overlap distribution inevitably implies the spin glass order indicated by the non-differentiability of the two-replica free energy or, equivalently, by a nonzero jump in the expectation value of the replica overlap.
It should be noted that nonzero $\qjump$ is, in general, not related to any symmetry breaking.  In this sense, Theorem~\ref{t:Grgen} has a slightly different nature than the original Griffiths' theorem.

We note that the inequality \rlb{Grqjqbr} is clearly not optimal since it reduces to $\qEA\ge(\qbr)^2/4$ for a $\bbZ_2$ symmetric model, where we proved $\qEA\ge\qbr$.
We expect from a heuristic argument that the optimal and general inequality is $\qjump\ge\qbr$.
See \rlb{qj>qbr} in section~\ref{s:dis}.
It is interesting to see if this conjectured inequality can be proved generally by improving our method.

\subsection{Spontaneous breakdown of replica permutation symmetry}
\label{s:LRSB}
We shall discuss the characterization of spin glass order as a spontaneous breakdown of the permutation symmetry of multiple replicas.
One may call such a phenomenon a literal RSB.
As far as we know, such a viewpoint was first introduced by Guerra \cite{Guerra2013}.
See also the remark at the end of section~3 of \cite{M} for a related observation in the two-replica system for the random energy model.

To this end, we consider the system of three replicas described by the Hamiltonian \rlb{H3}, where the replicas 1 and 2 are coupled by the coupling parameter $\la$, and 1 and 3 by $\la'$.
It is crucial here  to note that the model is exactly  symmetric under any permutation of three replicas when $\lambda=\lambda'=0$.
In what follows we shall set $\lambda'=-\lambda$, and regard $\lambda >0 $ as a symmetry breaking field
that detects possible spontaneous breakdown of the permutation symmetry.
Observe that the transposition of the replicas 2 and 3 leads to the symmetry $f^{(3)}(\beta,\la,\la')=f^{(3)}(\beta,\la',\la)$ for the corresponding three-replica free energy, as we noted in \rlb{fp}.
The symmetry implies that $f^{(3)}(\beta,\la,-\la)$ is even in $\la$, and its derivative at $\la=0$ should vanish if anything unusual does not take place.
This motivates us to define a new order parameter
\eq
\qrsb\coloneqq-\lim_{\la\dn0}\frac{\partial}{\partial\la}f^{(3)}(\beta,\la,-\la),
\lb{qrsb}
\en
where ``lrsb" stands for literal RSB. 
As in \rlb{qEAEE}, this is rewritten in terms of the expectation values in the three-replica system as
\eq
\qrsb=\lim_{\la\dn0}\limL\bbE\Bigl\{\sbkt{R_L^{1,2}}^{(3)}_{L,\beta,\la,-\la;\calJ_L,\bsb(\beta,\calJ_L)}
-\sbkt{R_L^{1,3}}^{(3)}_{L,\beta,\la,-\la;\calJ_L,\bsb'(\beta, \calJ_L)}\Bigr\},
\en
for any boundary configurations $\bsb(\cdot)$ and $\bsb'(\cdot)$. 
We stress that $\qrsb\ne0$ implies $\sbkt{R^{1,2}}\ne\sbkt{R^{1,3}}$, which can be interpreted as a literal RSB.
Here, one should recall that the transposition symmetry $2\leftrightarrow3$ is only infinitesimally broken in the limit $\la\dn0$.

Since the following theorem applies to a wide range of models, including long-range spin glass models such as the SK model \cite{SK} or the random energy model \cite{Derrida1980,Derrida1981}, we shall state and prove it as a standalone statement.
\begin{theorem}[Spontaneous breakdown of replica permutation symmetry]
\label{t:rsb}
Let the three-replica free energy $f^{(3)}(\beta,\la,\la')$ be a concave function of $(\la,\la')\in\bbR^2$ with symmetry $f^{(3)}(\beta,\la,\la')=f^{(3)}(\beta,\la',\la)$.
We also assume that
\eq
f^{(3)}(\beta,\la,0)=f^{(2)}(\beta,\la)+f(\beta),
\lb{f3f2}
\en
for some functions $f^{(2)}(\beta,\la)$ and $f(\beta)$ (which are the two-replica and one-replica free energies, respectively).
If we define $\qjump$ and $\qrsb$ by \rlb{qjump} and \rlb{qrsb}, respectively,  it holds that
\eq
\qrsb\ge2\qjump.
\lb{rsbjump}
\en
\end{theorem}

\par\noindent
{\em Proof}\/: Recall that concavity implies
\eq
f^{(3)}(\beta,\la,\la')\le f^{(3)}(\beta,0,0)+(g,g')\cdot(\la,\la'),
\en
for any $(\la,\la')\in\bbR^2$, where $(g,g')$ is a subgradient of $f^{(3)}(\beta,\la,\la')$ at $(\la,\la')=(0,0)$.
This implies
\eq
-\frac{f^{(3)}(\beta,\la,-\la)-f^{(3)}(\beta,0,0)}{\la}\ge-g+g',
\lb{flala}
\en
for any $\la>0$.
By setting
\eqg
g=\lim_{\la\dn0}\frac{\partial}{\partial\la}f^{(3)}(\beta,\la,0)=\lim_{\la\dn0}\frac{\partial}{\partial\la}f^{(2)}(\beta,\la),\\
g'=\lim_{\la'\up0}\frac{\partial}{\partial\la'}f^{(3)}(\beta,0,\la')=\lim_{\la\up0}\frac{\partial}{\partial\la}f^{(2)}(\beta,\la),
\eng
the inequality \rlb{flala} yields \rlb{rsbjump} if we let $\la\dn0$.~\qed

\medskip

Note that \rlb{f3f2} is nothing but the infinite-volume limit of \rlb{f3}.
We thus see that a nonzero jump in the derivative of the two-replica free energy at $\la=0$ inevitably implies a spontaneous breakdown of the replica permutation symmetry in the three-replica system.

It should be noted that a nonzero $\qrsb$ does not necessarily imply an intrinsic spin glass order.
It is easily seen that the standard ferromagnetic order implies $\qrsb >0$. See the discussion after  
(5.6) in section \ref{s:dis}.

As far as we know, the existence of a nonzero jump in the derivatives of the two-replica free energy
in models (with or without $\bbZ_2$ symmetry) that are free from ferromagnetic order has been rigorously established only in long-ranged models, namely, the random energy model (REM) \cite{Guerra2013}\footnote{%
To be precise, the Kronecker delta coupling $\delta(\bss^1,\bss^2)$ between the replicas was employed in \cite{Guerra2013}.
We shall discuss in Appendix~\ref{s:REM} the REM with the coupling $N^{-1}\sum \sigma^1_x \sigma^2_x$.
} and the SK model \cite{T1,T,Guerra2003}\footnote{%
See, in particular, pp.127--128 of \cite{T1} or pp.96--98 in volume I of \cite{T}. 
}.
We conclude that these models exhibit a literal RSB in the three-replica system, as was shown in \cite{Guerra2013} for the REM.

The present theorem is most meaningful in the models without $\bbZ_2$ symmetry discussed in section~\ref{s:nonZ2}.
By combining Theorem~\ref{t:rsb} with Theorem~\ref{t:Grgen}, we get the following.
\begin{corollary}[Broadening of the overlap implies literal replica symmetry breaking]
For the general EA model without $\bbZ_2$ symmetry, we have
\eq
\qrsb\ge\frac{(\qbr)^2}{2}.
\en
\end{corollary}
The corollary shows that the spin glass order characterized by the broadening of the overlap distribution, which is usually regarded as a sign of RSB, inevitably implies the presence of literal RSB.
This observation is interesting since the model without $\bbZ_2$ symmetry does not exhibit any spontaneous symmetry breaking by itself or in the two-replica system.

\section{Proofs}\label{s:proof}
\subsection{Proofs of Theorems~\protect\ref{t:GrZ2} and \protect\ref{t:Grgen}}
The following lemma is the core of our theory.
\begin{lemma}
\label{L:main}
For any boundary condition $\bsb(\cdot)$, it holds that
\eq
-\lim_{\la\dn0}\frac{\partial}{\partial\la}f^{(2)}(\beta,\la)\ge\limsupL\sqrt{\bbE\,\bbkt{(R_L^{1,2})^2}^{(2)}_{L,\beta,0;\calJ_L,\bsb(\beta,\calJ_L)}}.
\lb{main}
\en
\end{lemma}
We shall prove the lemma in the next subsection.

\medskip
\noindent
{\em Proof of Theorem~\ref{t:GrZ2} given Lemma~\ref{L:main}}\/:
It suffices to note that the left-hand side of \rlb{main} is  $\qEA$ in \rlb{qEAf} and the right-hand side with $\bsb=\bszero$ is $\qbr$ in \rlb{qbr1}.~\qed

\medskip
\noindent
{\em Proof of Theorem~\ref{t:Grgen} given Lemma~\ref{L:main}}\/:
Let $(L_i)_{i=1,2,\ldots}$ be a subsequence that attains the $\limsup$ in \rlb{qbr2}, which defines $\qbr$.
We rewrite \rlb{qbr2} as
\eq
(\qbr)^2=\lim_{i\up\infty}\Bigl\{\Bigl(
\sqrt{\bbE\,\bbkt{(R_{L_i}^{1,2})^2}^{(2)}_{i}}
-\Bigl|\bbE\,\bbkt{R_{L_i}^{1,2}}^{(2)}_{i}\Bigr|\Bigr)
\Bigl(
\sqrt{\bbE\,\bbkt{(R_{L_i}^{1,2})^2}^{(2)}_{i}}
+\Bigl|\bbE\,\bbkt{R_{L_i}^{1,2}}^{(2)}_{i}\Bigr|\Bigr)\Bigr\},
\lb{qbr3}
\en
where $\sbkt{\cdot}_i$ is the abbreviation of $\sbkt{\cdot}_{L_i,\beta,0;\calJ_L,\bszero}$.
Since the first factor is non-negative and the second does not exceed 2 for every $L_i$, we find 
\eq
\lim_{i\up\infty}\Bigl\{\sqrt{\bbE\,\bbkt{(R_{L_i}^{1,2})^2}^{(2)}_{i}}-\Bigl|\bbE\,\bbkt{R_{L_i}^{1,2}}^{(2)}_{i}\,\Bigr|\Bigr\}
\ge\frac{(\qbr)^2}{2},
\lb{qbr4}
\en
from which we conclude
\eq
\limsupL\sqrt{
\bbE\,\bbkt{(R_L^{1,2})^2}^{(2)}_{L,\beta,0;\calJ_L,\bszero}}
\ge
\liminfL\Bigl|\bbE\,\bbkt{R_L^{1,2}}^{(2)}_{L,\beta,0;\calJ_L,\bszero}\Bigr|+\frac{(\qbr)^2}{2}.
\lb{qbr5}
\en
Recalling that concavity of $f^{(2)}(\beta,\la)$ and \rlb{f2la} imply
\eq
\liminfL\bbE\,\bbkt{R_L^{1,2}}^{(2)}_{L,\beta,0;\calJ_L,\bszero}\ge-\lim_{\la\up0}\frac{\partial}{\partial\la}f^{(2)}(\beta,\la),
\lb{Rf}
\en
we get the desired \rlb{Grqjqbr} from \rlb{main} and \rlb{qbr5}.~\qed\\

The proof of Lemma~\protect\ref{L:main} given in the following subsection essentially relies on the short-range nature of the model. This means our proofs of Theorems~\ref{t:GrZ2} and ~\ref{t:Grgen} do not apply to long-range models.  However, the theorems are expected to be valid also for long-range models.  See section~\ref{s:dis}.

\subsection{Proof of Lemma~\protect\ref{L:main}}
\label{s:proofmain}
Our goal is to show that 
\eq
\qEA\ge\limsupL\sqrt{\bbE\,\bbkt{(R_L^{1,2})^2}^{(2)}_{L,\beta,0;\calJ_L,\bsb(\beta,\calJ_L)}},
\lb{main2}
\en
where $\qEA$ is the EA order parameter defined in \rlb{qEA1} for $\bbZ_2$ symmetric models.
Here, we use exactly the same definition for general models, where $\qEA$ no longer plays the role of an order parameter.
The existence of the limit in \rlb{qEA1} is proved at the end of the present section.

The desired \rlb{main} follows from \rlb{main2} along with $-\lim_{\la\dn0}\frac{\partial}{\partial\la}f^{(2)}(\beta,\la)\ge\qEA$, which, like \rlb{Rf}, is a simple and standard consequence of concavity.
In fact, the corresponding equality  $-\lim_{\la\dn0}\frac{\partial}{\partial\la}f^{(2)}(\beta,\la)=\qEA$ is also valid in models without $\bbZ_2$ symmetry, as was proved by van Enter and Griffiths \cite{vEG}.

\medskip

Let us prove \rlb{main2}.
The basic idea is to use a simple correlation inequality devised in \cite{HT} to study critical phenomena in random spin systems.

\begin{figure}
\begin{center}
{\includegraphics[width=7truecm]{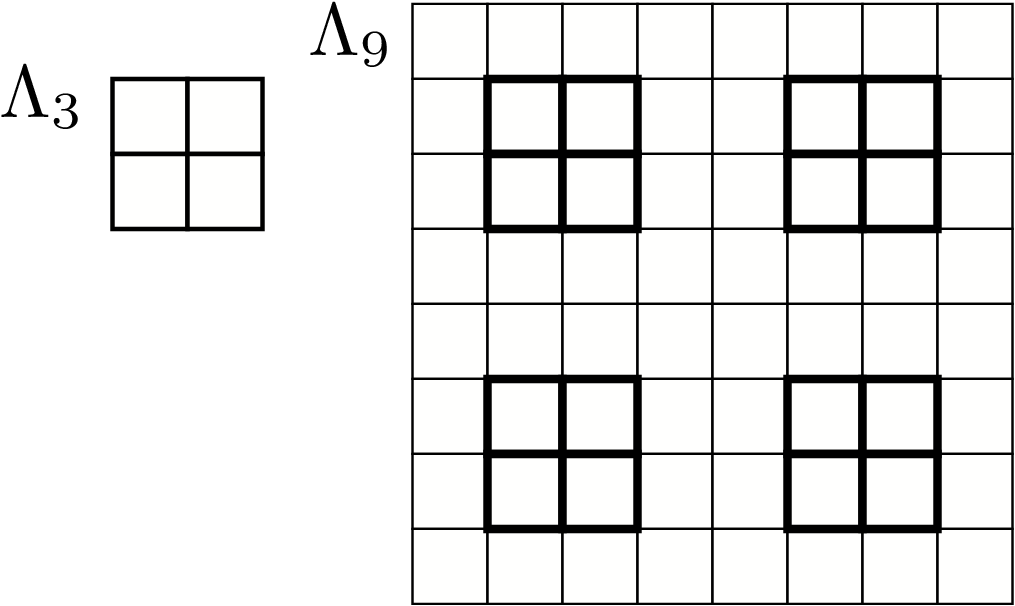}}
\caption[dummy]{
Four translated copies of $\La_3$ are embedded into $\La_9$.
This corresponds to the optimal choice \rlb{Kopt}.
}
\label{f:La39}
\end{center}
\end{figure}
Fix $L$, take $\ell$ that is much smaller than $L$, and let $\Lal$ be the $d$-dimensional $\ell\times\cdots\ell$ hypercubic lattice as in \rlb{LaL}.
We embed $K$ translated copies of $\Lal$ into $\LaL$ so that no two copies are closer than distance two.
More precisely, we denote by $\Lal^\kappa\subset\LaL$ with $\kappa=1,\ldots,K$ the translated copies of $\Lal$ and assume for any $\kappa\ne\kappa'$ that $|x-y|\ge2$ for all $x\in\Lal^\kappa$ and $y\in\Lal^{\kappa'}$.
For notational convenience, we also assume that copies of $\Lal$ do not touch the boundary of $\LaL$, i.e., $|x-u|\ge2$ for any $x\in\cup_{\kappa=1}^K\Lal^\kappa$ and $u\in\dLaL$.
See Figure~\ref{f:La39}.

Fix any $\kappa\ne\kappa'$ and abbreviate $\Lal^\kappa$ and $\Lal^{\kappa'}$ as $\Laa$ and $\Lab$, respectively.
We decompose any spin configuration $\bss=(\sigma_x)_{x\in\LaL}\in\calC_L$ as $\bss=(\bssa,\bssb,\bstau)$, where $\bssa=(\sigma_x)_{x\in\Laa}$, $\bssb=(\sigma_x)_{x\in\Lab}$, and $\bstau=(\tau_x)_{x\in\LaL\backslash(\Laa\cup\Lab)}$. Here we wrote $\tau_x=\sigma_x$ for later convenience.

Let us tentatively fix a boundary condition $\bsb$ and a random realization of $J_{x,y}$ and $h_x$, and decompose the Hamiltonian \rlb{RSHamil} as
\eq
H_L(\bss;\bsb)=H_\mathrm{a}(\bssa,\bstau)+H_\mathrm{b}(\bssb,\bstau)+\tilde{H}(\bstau,\bsb),
\lb{HHH}
\en
where, for $\alpha=\mathrm{a}, \mathrm{b}$, we set
\eq
H_\alpha(\bss_\alpha,\bstau)=- \sumtwo{\{x,y\}\in\calB_L}{{\rm s.t.}\,x,y\in\La_\alpha} J_{x,y}\,\sigma_x \sigma_y
-\sumtwo{\{x,y\}\in\calB_L}{{\rm s.t.}\,x\in\La_\alpha,\,y\in\LaL\backslash\La_\alpha}J_{x,y}\,\sigma_x\tau_y
- \sum_{x \in \La_\alpha}h_x\,\sigma_x.
\en
It is crucial that $\tilde{H}$ does not depend on $\bssa$ or $\bssb$.
We have tentatively dropped the $\calJ_L$ dependence for notational simplicity.

Take any $x\in\Laa$ and $y\in\Lab$.
We consider the standard expectation value of $\sigma_x\sigma_y$ as defined in \rlb{<F>}, and rewrite it as
\eqa
\sbkt{\sigma_x\sigma_y}_{L,\beta;\bsb}&=\frac{1}{Z_{L,\beta;\bsb}}\sum_{\bss\in\calC_L}\sigma_x\sigma_y\,e^{-\beta H_L(\bss;\bsb)}
\nl
&=\frac{1}{Z_{L,\beta;\bsb}}\sum_{\bstau}e^{-\beta\tilde{H}(\bstau;\bsb)}
\sum_{\bssa}\sigma_x\,e^{-\beta H_\mathrm{a}(\bssa,\bstau)}
\sum_{\bssb}\sigma_y\,e^{-\beta H_\mathrm{b}(\bssb,\bstau)}
\nl
&=\frac{1}{Z_{L,\beta;\bsb}}\sum_{\bstau}e^{-\beta\tilde{H}(\bstau;\bsb)}\,
Z_{\mathrm{a},\bstau}\,\sbkt{\sigma_x}_{\mathrm{a},\bstau}\,
Z_{\mathrm{b},\bstau}\,\sbkt{\sigma_y}_{\mathrm{b},\bstau}
\lb{sxsy1}
\ena
where, for $\alpha=\mathrm{a}, \mathrm{b}$, we defined
\eq
\sbkt{\cdots}_{\alpha,\bstau}=\frac{1}{Z_{\alpha,\bstau}}\sum_{\bss_\alpha}(\cdots)\,e^{-\beta H_\alpha(\bss_\alpha,\bstau)},\quad
Z_{\alpha,\bstau}=\sum_{\bss_\alpha}e^{-\beta H_\alpha(\bss_\alpha,\bstau)}.
\en
By defining
\eq
P(\bstau)=\frac{e^{-\beta\tilde{H}(\bstau;\bsb)}\,Z_{\mathrm{a},\bstau}\,Z_{\mathrm{b},\bstau}}{Z_{L,\beta;\bsb}},
\en
which satisfies $P(\bstau)\ge0$ and $\sum_{\bstau}P(\bstau)=1$, we can write \rlb{sxsy1} as
\eq
\sbkt{\sigma_x\sigma_y}_{L,\beta;\bsb}=\sum_{\bstau}P(\bstau)\,\sbkt{\sigma_x}_{\mathrm{a},\bstau}\,\sbkt{\sigma_y}_{\mathrm{b},\bstau}.
\en
We then observe that
\eqa
\sumtwo{x\in\Laa}{y\in\Lab}\bigl(\sbkt{\sigma_x\sigma_y}_{L,\beta;\bsb}\bigr)^2&=\sumtwo{x\in\Laa}{y\in\Lab}\sum_{\bstau,\bstau'}P(\bstau)\,P(\bstau')\,\sbkt{\sigma_x}_{\mathrm{a},\bstau}\,\sbkt{\sigma_x}_{\mathrm{a},\bstau'}\,\sbkt{\sigma_y}_{\mathrm{b},\bstau}\,\sbkt{\sigma_y}_{\mathrm{b},\bstau'}
\nl&\le\sumtwo{x\in\Laa}{y\in\Lab}\sum_{\bstau}P(\bstau)\,\bigl(\sbkt{\sigma_x}_{\mathrm{a},\bstau}\bigr)^2\,\bigl(\sbkt{\sigma_y}_{\mathrm{b},\bstau}\bigr)^2
\nl&=\sum_{\bstau}P(\bstau)\Bigl\{\sum_{x\in\Laa}\bigl(\sbkt{\sigma_x}_{\mathrm{a},\bstau}\bigr)^2\Bigr\}
\Bigl\{\sum_{y\in\Lab}\bigl(\sbkt{\sigma_y}_{\mathrm{b},\bstau}\bigr)^2\Bigr\},
\lb{sxsy2}
\ena
where we used the trivial inequality
\eq
\sbkt{\sigma_x}_{\mathrm{a},\bstau}\,\sbkt{\sigma_x}_{\mathrm{a},\bstau'}\,\sbkt{\sigma_y}_{\mathrm{b},\bstau}\,\sbkt{\sigma_y}_{\mathrm{b},\bstau'}
\le\frac{1}{2}\Bigl\{\bigl(\sbkt{\sigma_x}_{\mathrm{a},\bstau}\bigr)^2\,\bigl(\sbkt{\sigma_y}_{\mathrm{b},\bstau}\bigr)^2+\bigl(\sbkt{\sigma_x}_{\mathrm{a},\bstau'}\bigr)^2\,\bigl(\sbkt{\sigma_y}_{\mathrm{b},\bstau'}\bigr)^2\Bigr\},
\en
to get the second line.

Recall that (a part of) the configuration $\bstau$ plays the role of boundary condition for the expectation value $\sbkt{\cdots}_{\alpha,\bstau}$.
This means that the right-most hand of \rlb{sxsy2} may be upper-bounded as
\eq
\sumtwo{x\in\Laa}{y\in\Lab}\bigl(\sbkt{\sigma_x\sigma_y}_{L,\beta;\calJ_L,\bsb}\bigr)^2
\le\max_{\bsb'}\sum_{x\in\Laa}\bigl(\sbkt{\sigma_x}_{\mathrm{a};\calJ_L,\bsb'}\bigr)^2\,\max_{\bsb''}\sum_{x\in\Lab}\bigl(\sbkt{\sigma_x}_{\mathrm{b};\calJ_L,\bsb''}\bigr)^2.
\lb{sxsy3}
\en
Here $\sbkt{\cdots}_{\mathrm{a};\calJ_L,\bsb'}$ denotes the expectation value, exactly as in \rlb{<F>}, on the lattice $\Laa$ with boundary configuration $\bsb'$ and random interactions and fields determined by 
$\calJ_L$.

We now allow the boundary configuration $\bsb$ (for the larger lattice $\LaL$) to depend on $\calJ_L$.
By taking the random average of \rlb{sxsy3}, we get our main inequality
\eq
\bbE\sum_{x\in\Lal^\kappa,\,y\in\Lal^{\kappa'}}\bigl(\sbkt{\sigma_x\sigma_y}_{L,\beta;\calJ_L,\bsb(\beta,\calJ_L)}\bigr)^2
\le\ell^{2d}\,\bigl\{\qEA(\ell)\bigr\}^2,
\lb{sxsy4}
\en
for any $\kappa,\kappa'=1,\ldots,K$ with $\kappa\ne\kappa'$, where $\qEA(\ell)$ is defined in \rlb{qL}.
Here we noted that the two expectation values on the right-hand side of \rlb{sxsy3} are independent.

To complete the proof, we consider the decomposition of the summation
\eq
\sum_{x,y\in\LaL}(\cdots)=\mathop{\sum_{\kappa,\kappa'=1}^K}_{{\rm s.t.}\,\kappa\ne\kappa'}\sum_{x\in\Lal^\kappa,\,y\in\Lal^{\kappa'}}(\cdots)
+\sum_{(x,y)\in\mathcal{R}}(\cdots),
\lb{sumdec}
\en
where $\mathcal{R}$ is defined by this equation.
The set $\mathcal{R}$ contains pairs of $x$ and $y$ that belong to the same translated copy $\Lal^\kappa$, and pairs of $x$ and $y$ with at least one of them not belonging to any $\Lal^\kappa$.
It is crucial to us that the first sum is over $K(K-1)\,\ell^{2d}$ terms and the second sum is over $L^{2d}-K(K-1)\,\ell^{2d}$ terms.
We then find
\eq
\bbE\sum_{x,y\in\LaL}\bigl(\sbkt{\sigma_x\sigma_y}_{L,\beta;\calJ_L,\bsb(\beta,\calJ_L)}\bigr)^2
\le K(K-1)\,\ell^{2d}\,\bigl\{\qEA(\ell)\bigr\}^2+\bigl\{L^{2d}-K(K-1)\,\ell^{2d}\bigr\},
\lb{sxsy5}
\en
where we used \rlb{sxsy4} for the first sum in the right-hand side of \rlb{sumdec} and $\bigl(\sbkt{\sigma_x\sigma_y}_{L,\beta;\calJ_L,\bsb(\beta,\calJ_L)}\bigr)^2\le1$ for the second sum.
Note that the optimal choice of the number of translated copies is
\eq
K=\Bigl\lfloor\frac{L-1}{\ell+1}\Bigr\rfloor^d,
\lb{Kopt}
\en
which implies
\eq
\limL\frac{K}{L^d}=\frac{1}{(\ell+1)^d}.
\en
We then find from \rlb{sxsy5} that
\eqa
\limsupL\bbE\,\bbkt{(R_L^{1,2})^2}^{(2)}_{L,\beta,0;\calJ_L,\bsb(\beta,\calJ_L)}&=
\limsupL\frac{1}{L^{2d}}\,\bbE\sum_{x,y\in\LaL}\bigl(\sbkt{\sigma_x\sigma_y}_{L,\beta;\calJ_L,\bsb(\beta,\calJ_L)}\bigr)^2
\nl
&\le\Bigl(\frac{\ell}{\ell+1}\Bigr)^{2d}\,\{\qEA(\ell)\}^2+\Bigl\{1-\Bigl(\frac{\ell}{\ell+1}\Bigr)^{2d}\Bigr\},
\ena
for any $\ell$.
By taking $\liminf_{\ell\up\infty}$, we get 
\eq
\limsupL\bbE\,\bbkt{(R_L^{1,2})^2}^{(2)}_{L,\beta,0;\calJ_L,\bsb(\beta,\calJ_L)}\le\liminfL\{\qEA(L)\}^2,
\lb{R2qEA2}
\en
which is the desired \rlb{main2} provided that the limit $\qEA=\limL\qEA(L)$ in \rlb{qEA1} exists.

To show the existence of the limit, let $\bsb_\mathrm{max}(\beta,\calJ_L)$ be a boundary condition that attains the maximum in \rlb{qL}.
Then, using the nonnegativity of variance twice, we see 
\eqa
\bbE\,\bbkt{(R_L^{1,2})^2}^{(2)}_{L,\beta,0;\calJ_L,\bsb_\mathrm{max}(\beta,\calJ_L)}
&\ge
\bbE\,\bigl(\bbkt{R_L^{1,2}}^{(2)}_{L,\beta,0;\calJ_L,\bsb_\mathrm{max}(\beta,\calJ_L)}\bigr)^2
\nl&\ge
\bigl(\bbE\,\bbkt{R_L^{1,2}}^{(2)}_{L,\beta,0;\calJ_L,\bsb_\mathrm{max}(\beta,\calJ_L)}\bigr)^2
=\{\qEA(L)\}^2.
\lb{R2qEA2B}
\ena
By taking $\limsupL$ and using \rlb{R2qEA2}, we find
\eq
\limsupL\{\qEA(L)\}^2\le\liminfL\{\qEA(L)\}^2.
\en
Since $\qEA(L)\ge0$, we see that the limit $\limL\qEA(L)$ exists.

\bigskip
\noindent
{\em Remark}\/:  From \rlb{R2qEA2} and  \rlb{R2qEA2B}, we see
\eq
\limL\bbE\,\Bigl\{\bbkt{(R_L^{1,2})^2}^{(2)}_{L,\beta,0;\calJ_L,\bsb_\mathrm{max}(\beta,\calJ_L)}-\bigl(\bbkt{R_L^{1,2}}^{(2)}_{L,\beta,0;\calJ_L,\bsb_\mathrm{max}(\beta,\calJ_L)}\bigr)^2\Bigr\}=0,
\en
which means that the variance of the overlap vanishes in the equilibrium state with the boundary condition $\bsb_\mathrm{max}(\beta,\calJ_L)$, as was pointed out in \cite{I2,I4}. 
In other words, the broadening order parameter $\qbr$ defined by \rlb{qbr2} with this boundary condition is always zero.

\subsection{Remarks about boundary conditions}
\label{s:BC}
In the definitions \rlb{qbr1} and \rlb{qbr2} of the order parameter $\qbr$ for the broadening of the overlap distribution, we systematically used the open boundary condition only because it is most natural from a physical point of view.
Technically speaking, it can be replaced by any boundary condition.

To see this, let us first examine the proof of Lemma~\ref{L:main}.
Here the boundary condition $\bsb$ for the whole lattice $\LaL$ enters only in $\tilde{H}(\bstau,\bsb)$ in the decomposition \rlb{HHH}.
Therefore $\bsb$ plays no role in the resulting bound such as \rlb{sxsy3}.
This means that the quantity
\eq
\bbE\,\bbkt{(R_L^{1,2})^2}^{(2)}_{L,\beta,0;\calJ_L,\bszero}=\bbE\sum_{x,y\in\LaL}\bigl(\sbkt{\sigma_x\sigma_y}_{L,\beta;\calJ_L,\bszero}\bigr)^2,
\en
in \rlb{qbr1} or \rlb{qbr2} may be replaced by
\eq
\bbE\sum_{x,y\in\LaL}\bigl(\sbkt{\sigma_x\sigma_y}_{L,\beta;\calJ_L,\bsb_1(\calJ_L)}\bigr)\bigl(\sbkt{\sigma_x\sigma_y}_{L,\beta;\calJ_L,\bsb_2(\calJ_L)}\bigr),
\en
where $\bsb_1(\cdot)$ and $\bsb_2(\cdot)$ are arbitrary boundary conditions.
It is clear that one can employ the periodic boundary condition (which is also physically natural), although it does not fit into our notation.

To see how the second term in the square root in \rlb{qbr2} can be generalized, we recall that this term is bounded as \rlb{Rf} in the proof of Theorem~\ref{t:Grgen}.
But this is a very general bound that is valid for any boundary condition.
This means that the quantity $(\bbE\,\bbkt{R_L^{1,2}}^{(2)}_{L,\beta,0;\calJ_L,\bszero})^2$ in \rlb{qbr2} may be replaced by
\eq
\Bigl(\bbE\sum_{x\in\LaL}\sbkt{\sigma_x}_{L,\beta;\calJ_L,\bsb_3(\calJ_L)}\sbkt{\sigma_x}_{L,\beta;\calJ_L,\bsb_4(\calJ_L)}\Bigr)
\Bigl(\bbE\sum_{x\in\LaL}\sbkt{\sigma_x}_{L,\beta;\calJ_L,\bsb_5(\calJ_L)}\sbkt{\sigma_x}_{L,\beta;\calJ_L,\bsb_6(\calJ_L)}\Bigr),
\en
where $\bsb_3(\cdot),\ldots,\bsb_6(\cdot)$ are arbitrary boundary conditions.

\section{Discussion}
\label{s:dis}
In the present paper, we proved three theorems that can be regarded as extensions to spin glass models of
Griffiths' theorem for ferromagnetic spin models \cite{Gff}.
Although we here only treated the Ising-type model with nearest-neighbor interactions, the theorems can readily be extended to any classical spin system with short-range interactions. 
It is only essential that the spin is bounded and the interaction and the magnetic field are stochastically translation invariant.
On the other hand, we are not able to extend Theorems~\ref{t:GrZ2} and \ref{t:Grgen} to quantum spin models.
We believe the difficulty is essential since the non-locality of quantum systems inhibits us from using the locality argument that led to the essential bound \rlb{sxsy3}.

In our theory, the most basic characterization of spin glass order is given by the broadening order parameter $\qbr$ defined by \rlb{qbr2}, which reduces to \rlb{qbr1} in the special case.
The order parameter $\qbr$ detects possible broadening of the probability distribution of the replica overlap $R^{1,2}$ in the equilibrium state with the open boundary condition.

In the standard EA model without a magnetic field, which has a global $\bbZ_2$ symmetry, it is standard to classify the equilibrium phases by the spontaneous magnetization $\mu_\mathrm{SM}$ defined as in \rlb{muSM} and the EA order parameter $\qEA$ \cite{EA}. 
Here, we followed van Enter and Griffiths \cite{vEG} and defined $\qEA$ by \rlb{qEA1}.
The spontaneous magnetization $\mu_\mathrm{SM}$ detects a spontaneous breaking of the $\bbZ_2$ symmetry of a single system while the EA order parameter $\qEA$ detects that of the $\bbZ_2$ symmetry \rlb{Z2} of the two-replica system.
Since the former implies the latter, that $\mu_\mathrm{SM}>0$ implies that $\qEA>0$, but not vice versa.
Consequently the paramagnetic phase is characterized by  $\mu_\mathrm{SM}=\qEA=0$, the spin glass phase by $\mu_\mathrm{SM}=0$, $\qEA\ne0$, and the ferromagnetic phase by $\mu_\mathrm{SM}\ne0$, $\qEA\ne0$.
In this case, our first theorem, Theorem~\ref{t:GrZ2}, shows that
\eq
\qEA\ge\qbr,
\lb{qEAqbr}
\en
where the broadening order parameter $\qbr$ is defined by \rlb{qbr1}.
We thus see that $\qbr>0$ and $\mu_\mathrm{SM}=0$ is a sufficient condition for the spin glass phase.
We note that the bound \rlb{qEAqbr} is valid if one replaces the open boundary condition in the definition \rlb{qbr1} by any other boundary condition, even one depending on $\beta$ and $\calJ_L$.

Our theory is probably most meaningful in the spin glass models without a global $\bbZ_2$ symmetry, such as the EA model with a nonzero (random or non-random) magnetic field.
Note that spin glass order in such models cannot be related to spontaneous symmetry breaking.
To detect possible spin glass order, we employed the jump order parameter $\qjump$ defined as \rlb{qjump}, following van Enter and Griffiths \cite{vEG}.
It represents the discontinuity in the derivative of the two-replica free energy with respect to the replica coupling parameter $\la$.
We stress that it is traditional in statistical mechanics (and thermodynamics) to characterize a phase transition in terms of the singularity in the free energy.
The discontinuity $\qjump >0$ has been established only in the long-range spin glass models, namely, for the SK model in the region of AT instability \cite{AT,T1,T} and the random energy model at low temperatures \cite{Guerra2013,M}.   
As we discussed at the beginning of section~\ref{s:nonZ2}, whether a short-range model without a $\bbZ_2$ symmetry exhibit spin glass phase transition is still controversial.
Our second theorem, Theorem~\ref{t:Grgen}, relates the broadening order parameter and the jump order parameter as
\eq
\qjump\geq \frac{(\qbr)^2}{4}.
\lb{qjqbr}
\en
Note that while $\qjump$ defined in terms of the free energy is insensitive to the boundary condition, $\qbr$ defined in \rlb{qbr2} as the standard deviation of $R^{1,2}$ explicitly depends on the choice of the boundary condition.
See, in particular, the remark at the end of section~\ref{s:proofmain}.
We here employed the open boundary condition (to define $\qbr$) since it is expected that the fluctuation of $R^{1,2}$ is large in this boundary condition, but we should note the bound \rlb{qjqbr} is valid for any boundary conditions.
See section~\ref{s:BC}.
This suggests that the inequality \rlb{qjqbr} provides a strategy for proving the absence, rather than the presence, of a spin glass phase transition in these models; if one proves that $\qjump=0$ then it implies that $\qbr=0$ for any choice of boundary conditions.
We should note, however, that this strategy does not work (at least) for the random field Ising model.
In \cite{Chatterjee2015}, Chatterjee established the absence of replica symmetry breaking in this model by proving $\qbr=0$ for the open boundary condition. 
It can be shown, on the other hand, that $\qjump>0$ in the same model in $d\ge3$ provided that the random magnetic field is sufficiently small and the temperature is sufficiently low.
Thus the converse of Griffiths' theorem is not valid in this case.
See Appendix~\ref{s:RFIM} for more details.

Let us note here that statements like Theorems~\ref{t:GrZ2} and \ref{t:Grgen} can be derived (non-rigorously) from a heuristic argument based on the probability distribution $P_\beta(q)$ of the replica overlap $R^{1,2}$.
This means that these inequalities (and the conjectured stronger inequality \rlb{qj>qbr}) should also hold in long-range spin glass models.
(We shall check this explicitly for the random energy model in Appendix~\ref{s:REM}. See \rlb{REMqbr} and \rlb{REMqjump}.)
Let us assume that the limit $P_q(\beta)=\limL P_{L,\beta}(q)$ exists and defines a well-behaved probability distribution supported on the interval $[\qmin,\qmax]$, where $P_{L,\beta}(q)$ is defined in \rlb{PLq}.
We then see from \rlb{qbr2} and the general upper bound for standard deviation that
\eq
\qbr=\sqrt{\int_{\qmin}^{\qmax}dq\,q^2\,P_\beta(q)-\Bigl\{\int_{\qmin}^{\qmax}dq\,q\,P_\beta(q)\Bigr\}^2}\le\frac{\qmax-\qmin}{2}.
\lb{qbrh}
\en
Note that the equality holds if and only if the distribution has two equal peaks at $\qmin$ and $\qmax$.
We also see from \rlb{H2} and \rlb{F2} that
\eqa
\lim_{\la\dn0}\limL\bbE\sbkt{R^{1,2}}_{L,\beta,\la;\calJ_L,\bszero}
&=\lim_{\la\dn0}\limL\frac{\int_{-1}^1dq\,q\,e^{\beta\la L^d q}\,P_{L,\beta}(q)}{\int_{-1}^1dq\,e^{\beta\la L^d q}\,P_{L,\beta}(q)}
\nl&=\lim_{\la\dn0}\limL\frac{\int_{\qmin}^{\qmax}dq\,q\,e^{\beta\la L^d q}\,P_\beta(q)}{\int_{\qmin}^{\qmax}dq\,e^{\beta\la L^d q}\,P_\beta(q)}=\qmax.
\ena
Here, the first equality is exact, while the second equality follows by assuming $P_{L,\beta}(q)$ converges to the limit $P_\beta(q)$ sufficiently quickly.
Since we similarly have $\lim_{\la\up0}\limL\bbE\sbkt{R^{1,2}}_{L,\beta,\la;\calJ_L,\bszero}=\qmin$, we find from \rlb{GrE} that $\qjump=(\qmax-\qmin)/2$.
Combined with \rlb{qbrh}, this observation suggests that we should generally have
\eq
\qjump\ge\qbr,
\lb{qj>qbr}
\en
which is strictly stronger than our inequality \rlb{Grqjqbr}.
For a $\bbZ_2$ symmetric model, this reduces to $\qEA\ge\qbr$, which is our \rlb{GrZ2}.

Our third theorem, Theorem~\ref{t:rsb}, has a slightly different character.
It applies to a wide class of models, including the short-range and the long-range spin glass models, and states that
\eq
\qrsb\geq 2\qjump,
\en
where the new order parameter $\qrsb$ detects spontaneous breaking of replica permutation symmetry in the three-replica system.
We have thus established that the SK model, where $\qjump>0$ is known, exhibits ``literal replica symmetry breaking'', extending Guerra's observation \cite{Guerra2013} on the random energy model.
As we noted above, the spin glass order, $\qjump>0$, in a model without a $\bbZ_2$ symmetry is not related to any spontaneous symmetry breaking.
It is interesting that the same order implies the ``replica symmetry breaking'' in the most literal sense of the terminology.

\begin{figure}
\begin{center}
{\includegraphics[width=11truecm]{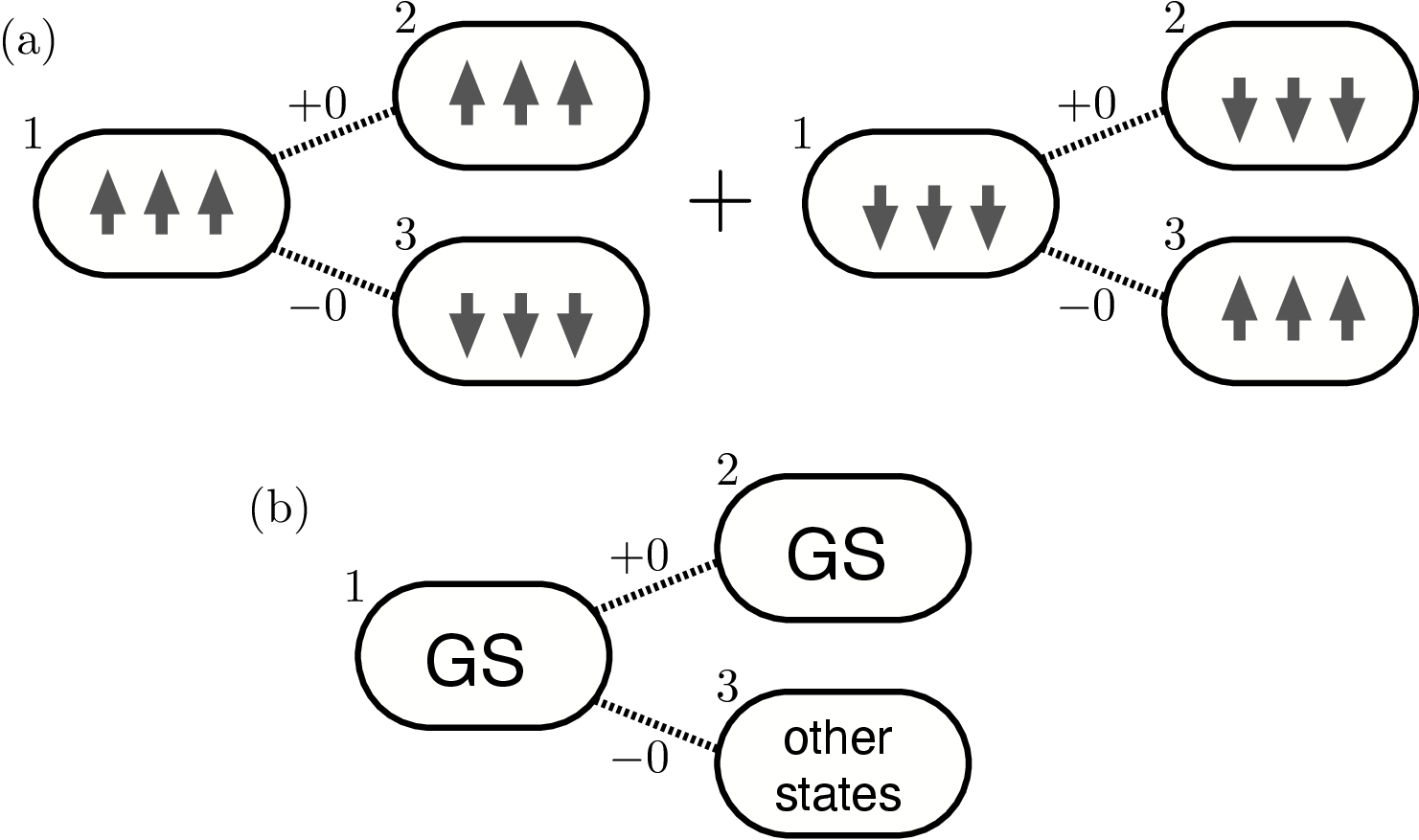}}
\caption[dummy]{
Schematic pictures of the low-temperature equilibrium states in the three-replica systems.
The rounded rectangles labeled as 1, 2, and 3 indicate replicas.
The dotted lines with $+0$ and $-0$ indicate coupling between the replicas with replica coupling parameters $\la\dn+0$ and $\la\up-0$, respectively.
(a)~The replicated ferromagnetic Ising model in two or higher dimensions at low temperatures.
The equilibrium state is the equal mixture of two states in which most spins in each replica are aligned with each other.
The spins in replicas 1 and 2 are pointing in the same direction, while the spins in replica 3 in the opposite direction.
Here, the replica permutation symmetry is broken, but it simply reflects the $\bbZ_2$ symmetry breaking that takes place in the single (non-replicated) system.
(b)~The duplicated random energy model at low temperatures.
The equilibrium state is a pure state in which the replicas 1 and 2 are in the ground state, and the replica 3 is in the mixture of all other spin configurations.
Here, the breaking of the replica permutation symmetry is the only symmetry breaking exhibited by the system and provides an intrinsic characterization of the spin glass order.
}
\label{f:RSB}
\end{center}
\end{figure}

We should note, however, that the order parameter $\qrsb$ does not always provide us with an intrinsic measure of the spin glass order.
In the ferromagnetic Ising model without any disorder, for example, the ferromagnetic order $\mu_\mathrm{SM}>0$ implies $\qbr>0$, $\qjump>0$, and $\qrsb>0$.
Here, the low-temperature equilibrium state of the three-replica system in the limit $\la^{1,2}=-\la^{1,3}\dn0$ (see \rlb{qrsb}) is the equal mixture of two states in which the spins in the replicas 1 and 2 are mostly pointing in the same direction and the spins in the replica 3 in the opposite direction.
See Figure~\ref{f:RSB}~(a).
In this case, the literal replica symmetry breaking indicated by $\qbr>0$ simply reflects the well-understood spontaneous breakdown of the $\bbZ_2$ symmetry in the original Ising model.
It is instructive to compare the situation with the random energy model \cite{BovierBook,Derrida1980,Derrida1981,DorlasWedagedera2000,Guerra2013,MM,M}
The low-temperature equilibrium state of the random energy model consists of a single spin configuration (called the ground state) whose weight is $1-\betac/\beta$ and all other spin configurations whose total weight is $\betac/\beta$.
Then, in the corresponding three-replica equilibrium state in the limit $\la^{1,2}=-\la^{1,3}\dn0$, the replicas 1 and 2 should be in the ground state while the replica 3  in the mixture of all other spin configurations.
See Figure~\ref{f:RSB}~(b).
Thus, the three-replica system exhibits spontaneous symmetry breaking of the replica permutation symmetry, while the corresponding single system does not break any symmetry.
We can say that, in this case, the literal replica symmetry breaking $\qrsb>0$ gives an intrinsic characterization of the spin glass order.
We conjecture that a somewhat similar picture applies to the low-temperature states of the SK model and also to the EA model under a magnetic field, provided that the latter is in the spin glass phase.

\appendix
\section{Random field Ising model}
\label{s:RFIM}
In section~\ref{s:dis} above, we argued that the jump order parameter $\qjump$ is strictly positive in the random field Ising model in $d\ge3$ provided that the magnetic field is sufficiently small and the temperature is sufficiently low, while Chatterjee's result $\qbr=0$ is always valid for this model.
We supply some details here.

The random field Ising model is defined within the general setting in section~\ref{s:single} by taking $J_{x,y}=1$ for all $x,y\in\bbZ^d$ such that $|x-y|=1$ and assuming that $h_x$ is drawn from the mean zero Gaussian distribution with variance $\ep^2$ (where we regard $\ep>0$).
It was proved by Bricmont and Kupiainen that the model exhibits ferromagnetic order in dimensions three or higher at sufficiently low temperatures when the random magnetic field is sufficiently weak \cite{BricmontKupiainen1988}.
Our assertion that $\qjump>0$ is based on this important result.
Let us prove this main claim.

\medskip\noindent{\em Proof of $\qjump>0$}\/:
Another important ingredient of the proof is the following general expression of $\qjump$ due to van Enter and Griffiths \cite{vEG}.
If $\rho_\alpha(\cdot)$ denotes an infinite-volume equilibrium state\footnote{%
We denote by $\rho_\alpha(F)$ the expectation value of any function $F(\bss)$.
} (for a given random interaction or field) labeled by $\alpha$, the jump order parameter is expressed as
\eq
\qjump=\frac{1}{2}\limL\Bigl[\max_\alpha L^{-d}\sum_{x\in\LaL}\bigl\{\rho_\alpha(\sigma_x)\bigr\}^2
-\min_{\alpha,\beta}L^{-d}\sum_{x\in\LaL}\rho_\alpha(\sigma_x)\rho_\beta(\sigma_x)\Bigr].
\lb{qjEG}
\en
See section~V of \cite{vEG}.

Let $\rho_+(\cdot)$ and $\rho_-(\cdot)$ be the states obtained as the infinite-volume limits of the finite-volume equilibrium states with the plus and minus boundary conditions, respectively.
(It is well known that the limiting states are well-defined.)
We then observe that
\eq
\limL\max_\alpha L^{-d}\sum_{x\in\LaL}\bigl\{\rho_\alpha(\sigma_x)\bigr\}^2\ge
\limL L^{-d}\sum_{x\in\LaL}\bigl\{\rho_+(\sigma_x)\bigr\}^2=\bbE\bigl\{\rho_+(\sigma_x)\bigr\}^2,
\lb{rr>}
\en
where the final identity follows from the self-averaging property and is valid for any $x$.
We similarly have
\eq
\limL\min_{\alpha,\beta}L^{-d}\sum_{x\in\LaL}\rho_\alpha(\sigma_x)\rho_\beta(\sigma_x)
\le\limL L^{-d}\sum_{x\in\LaL}\rho_+(\sigma_x)\rho_-(\sigma_x)
=\bbE\bigl\{\rho_+(\sigma_x)\rho_-(\sigma_x)\bigr\}.
\lb{rr<}
\en

For $d\ge3$, sufficiently large $\beta$, and sufficiently small $\ep$, Bricmont and Kupiainen \cite{BricmontKupiainen1988} proved that
\eq
\rho_+(\sigma_x)\ge1-\kappa,
\lb{sigma+}
\en
with probability not less than $1-\eta$, where $\kappa=e^{-O(\beta)}$ and $\eta=e^{-O(\ep^{-2})}$ are constants.
By symmetry, we also have
\eq
\rho_-(\sigma_x)\le-(1-\kappa),
\lb{sigma-}
\en
with probability not less than $1-\eta$.
From \rlb{sigma+} and \rlb{sigma-}, we see
\eqg
\bbE\bigl\{\rho_+(\sigma_x)\bigr\}^2\ge(1-\kappa)^2(1-\eta)-\eta,\\
\bbE\bigl\{\rho_+(\sigma_x)\rho_-(\sigma_x)\bigr\}\le-(1-\kappa)^2(1-2\eta)+2\eta.
\eng
We thus conclude from \rlb{qjEG}, \rlb{rr>}, and  \rlb{rr<} that
\eq
\qjump\ge(1-\kappa)^2(1-\tfrac{3}{2}\eta)-\tfrac{3}{2}\eta,
\en
where the right-hand side is positive for sufficiently large $\beta$ and sufficiently small $\ep$.~\qed

\medskip
It should be obvious that the positivity of $\qjump$ is a direct consequence of the ferromagnetic order in the equilibrium states.
One might then wonder why the random field Ising model has vanishing $\qbr$ while the ferromagnetic Ising model (without a magnetic field) has nonzero $\qbr$.
This is related to an interesting property of disordered models, as we shall explain below.

Consider the ferromagnetic Ising model with $h_x=0$ for all $x$, and suppose that the equilibrium state has nonzero spontaneous magnetization $\mu_{\rm SM}\coloneqq\rho_+(\sigma_x)>0$.
In this case the equilibrium state for large $L$ realized under the open boundary condition is well-approximated by the equal mixture of the plus and the minus states, i.e., $\sbkt{\cdot}_{L,\beta;\bszero}\simeq\{\rho_+(\cdot)+\rho_-(\cdot)\}/2$.
One easily finds that such a state has $\qbr=(\mu_{\rm SM})^2$.
Let us note here that the exact equal mixture is a manifestation of the $\bbZ_2$ symmetry of the model.

In the random field Ising model, where the $\bbZ_2$ symmetry is broken by the magnetic field, on the other hand, the equilibrium state under the open boundary condition is expected to be well approximated by either $\rho_+(\cdot)$ or $\rho_-(\cdot)$, not by a mixture. Equivalently, in the language of metastate developed in \cite{NewmanStein1997,NewmanStein2002}, the open-b.c. metastate is dispersive and supported on the two states $\rho_+(\cdot)$ and $\rho_-(\cdot)$.
This claim, although not yet proved rigorously, is physically plausible since the sum $\sum_{x\in\LaL}h_x$ of the random field is typically of order $\ep\,L^{d/2}$.
It is then likely that either the plus or minus state is preferred\footnote{%
It is crucial here that we employ the open boundary condition.
If we use, say, the plus boundary condition, then there are $O(L^{d-1})$ boundary spins, which energetically overshadow the bias from the random field.
} depending on the sign of the sum $\sum_{x\in\LaL}h_x$.
See \cite{IacobelliKulske,vEGNS} for closely related rigorous results.
If the equilibrium state under the open boundary condition is approximated by $\rho_+(\cdot)$ or $\rho_-(\cdot)$, then the corresponding $\qbr$ is vanishing since the state has (almost) non-fluctuating total magnetization.
This is consistent with the result of Chatterjee \cite{Chatterjee2015}.
Let us finally remark that Chatterjee \cite{Chatterjee2023B} recently proposed an interesting variation of the random field Ising model in which the equilibrium state under the open boundary condition is approximated by a mixture of  $\rho_+(\cdot)$ and $\rho_-(\cdot)$ with certain weights 
depending on the realization of the random field.

\section{Random energy model}
\label{s:REM}
In \cite{Guerra2013}, Guerra studied the replicated version of the random energy model (REM) with explicit replica coupling given by the Kronecker delta $\delta(\bss^1,\bss^2)$ and studied the discontinuity in the expectation value of the overlap and the breakdown of the permutation symmetry of three replicas.
Physically speaking, it is expected that the Kronecker delta coupling is qualitatively not very different from the coupling $N^{-1}\sum \sigma^1_x \sigma^2_x$ in \rlb{H2}, which we employ throughout the paper.
For completeness, however, here we shall discuss the non-differentiability of the two-replica free energy in the REM when the standard coupling in the present paper is employed.
After quickly reviewing basic materials for the REM in section~\ref{s:REMreview}, we present an exact (but not rigorous) calculation of the two-replica free energy in section~\ref{s:REMf}.
In section~\ref{s:REMp}, we give a rigorous proof of the non-differentiability.

\subsection{The model and some known results}
\label{s:REMreview}
Here we define the REM, fix some notations (so that to be consistent with the main text), and briefly discuss known important results.
For more details, see the literature, e.g., \cite{BovierBook,Derrida1980,Derrida1981,DorlasWedagedera2000,Guerra2013,MM,M}.

The model is specified not by a lattice but by a positive integer $N$, indicating the number of spins.
A spin configuration is $\bss=(\sigma_1,\ldots,\sigma_N)\in\calC_N$ with $\sigma_j=\pm1$.
For each $N$, we let $E_{\bss}$ with $\bss\in\calC_N$ be the independent Gaussian random variable with mean zero and variance $N/2$.
For a single system, the Hamiltonian is nothing but the random energy, i.e., $H_N(\bss)=E_{\bss}$.
The partition function and the specific free energy in the infinite-volume are given by
\eq
Z_N(\beta)=\sum_{\bss\in\calC_N}e^{-\beta E_{\bss}},
\lb{REMZ1}
\en
and
\eq
f(\beta)=-\limN\frac{1}{\beta N}\bbE\log Z_N(\beta),
\lb{RSBf1}
\en
where $\bbE$ denotes the average over the distribution of $E_{\bss}$.
Note that we did not make explicit the dependence of $Z_N(\beta)$ on $(E_{\bss})_{\bss\in\calC_N}$.

We also consider the corresponding two-replica system with the Hamiltonian
\eq
H_N^{(2)}(\bss^1,\bss^2;\la)=E_{\bss^1}+E_{\bss^2}-\la\,\bss^1\cdot\bss^2,
\lb{REMH2}
\en
where
\eq
\bss^1\cdot\bss^2=\sum_{j=1}^N\sigma_j^1\sigma_j^2=N\,R^{1,2}(\bss^1,\bss^2). 
\en
Since we only discuss the two-replica system, we shall abbreviate $R^{1,2}$ as $R$ in what follows.
The corresponding partition function, the expectation value of a function $F(\bss^1,\bss^2)$, and the infinite-volume free energy are given by
\eqg
Z^{(2)}_N(\beta,\la)=\sum_{\bss^1,\bss^2\in\calC_N}e^{-\beta H_N^{(2)}(\bss^1,\bss^2;\la)},
\lb{REMZ2}\\
\sbkt{F}_{N,\beta,\la}^{(2)}=\frac{1}{Z^{(2)}_N(\beta,\la)}\sum_{\bss^1,\bss^2\in\calC_N}F(\bss^1,\bss^2)\,e^{-\beta H_N^{(2)}(\bss^1,\bss^2;\la)},
\lb{REM<F>}
\eng
and
\eq
f^{(2)}(\beta,\la)=-\limN\frac{1}{\beta N}\bbE\log Z^{(2)}_N(\beta,\la).
\lb{REMf2}
\en

It is known that the model exhibits a spin glass phase transition at the critical point
\eq
\betac=2\sqrt{\log2},
\lb{REMbetac}
\en
which is visible in the exact solution of the free energy:
\eq
f(\beta)=\begin{cases}
-\dfrac{1}{\beta}\log 2-\dfrac{\beta}{4},&\beta\le\betac;\\
-\sqrt{\log2},&\beta\ge\betac.
\end{cases}
\lb{REMfsol}
\en
The distribution of the overlap defined as in \rlb{PLq} behaves as
\eq
P_\beta(q)=\limN P_{N,\beta}(q)=
\begin{cases}
\delta(q),&\beta\le\betac;\\
\Bigl(1-\dfrac{\betac}{\beta}\Bigr)\,\delta(q-1)+\dfrac{\betac}{\beta}\,\delta(q),&\beta\ge\betac.
\end{cases}
\lb{REMPq}
\en
It is clearly seen that the distribution broadens in the spin glass phase with $\beta>\betac$.
Correspondingly, the expectation value of the overlap (without replica coupling) is given by
\eq
\limN\bbE\bbkt{R}^{(2)}_{N,\beta,0}=
\begin{cases}
0,&\beta\le\betac;\\[1pt]
1-\dfrac{\betac}{\beta},&\beta\ge\betac,
\end{cases}
\lb{REMR}
\en
and the broadening order parameter corresponding to \rlb{qbr2} by
\eq
\qbr\coloneqq
\limN\sqrt{\bbE\bbkt{R^2}^{(2)}_{N,\beta,0}-\Bigl(\bbE\bbkt{R}^{(2)}_{N,\beta,0}\Bigr)^2}
=
\begin{cases}
0,&\beta\le\betac;\\[2pt]
\sqrt{\dfrac{\betac}{\beta}\biggl(1-\dfrac{\betac}{\beta}\biggr)},&\beta\ge\betac.
\end{cases}
\lb{REMqbr}
\en

\subsection{Exact evaluation of the two-replica free energy}
\label{s:REMf}
Let us first demonstrate the non-differentiability of $f^{(2)}(\beta,\la)$ by explicitly evaluating it.
We believe that the following evaluation of $f^{(2)}(\beta,\la)$ is exact, but do not try to present a rigorous derivation.
We still do not know how difficult it would be to make our argument fully rigorous.

We first separate the summation in the two-replica partition function \rlb{REMZ2} to those with $\bss^1=\bss^2$ and with $\bss^1\ne\bss^2$ as
\eqa
Z^{(2)}_N(\beta,\la)&=\sum_{\bss^1,\bss^2\in\calC_N}
e^{-\beta E_{\bss^1}-\beta E_{\bss^2}+\beta\la\,\bss^1\cdot\bss^2}
\nl&=e^{\beta\la N}\sum_{\bss\in\calC_N}e^{-2\beta E_{\bss}}+\sum_{\bss^1\in\calC_N}e^{-\beta E_{\bss^1}}\sumtwo{\bss^2\in\calC_N}{\text{s.t.}\,\bss^2\ne\bss^1}e^{-\beta E_{\bss^2}+\beta\la\bss^1\cdot\bss^2}
\nl&=
e^{\beta\la N} Z_N(2\beta)+\sum_{\bss^1\in\calC_N}e^{-\beta E_{\bss^1}}\tZ_N(\bss^1;\beta,\la),
\lb{REMZ2A}
\ena
where we defined $\tZ_N(\bss^1;\beta,\la)$ in the final line.
By making the change of variable $\bss^2\to\bss$ with $\sigma_j=\sigma^1_j\sigma_j^2$ (for a fixed $\bss^1$) we see
\eq
\tZ_N(\bss^1;\beta,\la)=\sumtwo{\bss^2\in\calC_N}{\text{s.t.}\,\bss^2\ne\bss^1}e^{-\beta E_{\bss^2}+\beta\la\bss^1\cdot\bss^2}
=\sumtwo{\bss\in\calC_N}{\text{s.t.}\,\bss\ne(1,\ldots,1)}e^{-\beta E_{\bss}+\beta\la M(\bss)},
\en 
where $M(\bss)=\sum_{j=1}^N\sigma_j$ is the total magnetization.
Recalling that $E_{\bss}$ is a random quantity and that $\bss$ and $\bss^2$ are in one-to-one correspondence, we expect for sufficiently large $N$ the right-hand side may be replaced by
\eq
\tZ_N(\beta,\la)=\sumtwo{\bss\in\calC_N}{\text{s.t.}\,\bss\ne(1,\ldots,1)}e^{-\beta E_{\bss}+\beta\la M(\bss)},
\en
which is independent of $\bss^1$.
If we accept this substitution, we have
\eq
Z^{(2)}_N(\beta,\la)\simeq
e^{\beta\la N} Z_N(2\beta)+Z_N(\beta)\,\tZ_N(\beta,\la).
\lb{REMZ2main}
\en
We here note that 
\eq
\sum_{\bss\in\calC_N}e^{-\beta E_{\bss}+\beta\la M(\bss)}=\tZ_N(\beta,\la)+e^{\beta\la N}e^{-\beta E_{(1,\ldots,1)}},
\en
is nothing but the partition function of the REM under magnetic filed with the Hamiltonian $\tilde{H}(\bss)=E_{\bss}-\la M(\bss)$.
The model was discussed in \cite{Derrida1980,Derrida1981} and analyzed rigorously in \cite{DorlasWedagedera2000}.
It then suffices to substitute the known asymptotic behaviors of $Z_N(\beta)$ and $\tZ_N(\beta,\la)$ and then extract the term that dominates the right-hand side of \rlb{REMZ2main} for large $N$.
See, e.g., \cite{Guerra2013} for the second procedure.

By using \rlb{REMf2} and \rlb{REMfsol} for the asymptotic behavior of $Z_N(\beta)$ and the results proved and summarized in sections~3 and 4 of \cite{DorlasWedagedera2000} for the asymptotic behavior of $\tZ_N(\beta,\la)$, we arrive at the estimate
\eq
f^{(2)}(\beta,\la)=-\frac{1}{\beta}\max\{a_1(\beta,\la),a_2(\beta,\la)\},
\lb{maxa1a2}
\en
where we defined
\eq
a_1(\beta,\la)=
\begin{cases}
 \beta^2+\log2+\beta\la,&\beta\le\betac/2;\\
2\beta \sqrt{\log2}+\beta\la,&\beta>\betac/2,
\end{cases}
\lb{a1}
\en
and
\eq
a_2(\beta,\la)=
\begin{cases}
\dfrac{\beta^2}{2}+\log2+s_0(\tanh\beta\la)+\beta\la\tanh\beta\la,&\beta\le \bar\betac(\la);\\[3mm]
\dfrac{\beta^2}{4}+\log2+\beta \sqrt{s_0(\rho(\la))}+\beta\la\,\rho(\la),&\bar\betac(\la)<\beta < \betac;\\[3mm]
\beta \sqrt{\log2}+\beta\sqrt{s_0(\rho(\la))}+\beta\la\,\rho(\la),&\beta \ge\betac,
\end{cases}
\lb{a2}
\en
with the binary entropy
\eq
s_0(r)\coloneqq-\frac{1}{2}\Bigl\{(1+r) \log\frac{1+r}{2} + (1-r) \log\frac{1-r}{2} \Bigr\}.
\en
Here, $\bar\betac(\la)$ is the unique solution of  
\eq
  \bar\betac(\la) = 2 \sqrt{s_0(\tanh(\bar\betac(\la)\la))}.
\en
Note that $\bar\betac(\la)$ is decreasing in $|\lambda|$ and satisfies $\bar\betac(0)=\betac$ and  $\bar \beta_c(\la) \to 0$ as $\la \to \pm \infty$. 
In addition, $\rho(\la)$ is the unique solution of the equation
\eq
  \rho(\la) = \tanh\left( 2 \lambda \sqrt{s_0(\rho(\la))} \right),
\en
and has the same order of $\la$ as
\eq
  \rho(\la) = 2 \sqrt{\log 2}\,  \la + O(\la^3).
\en
The explicit expressions \rlb{a1} and \rlb{a2} allow us to evaluate the right-hand side of \rlb{maxa1a2} for sufficiently small $|\la|$. The result is 
\eq
-\beta f^{(2)}(\beta, \la)=\dfrac{\beta^2}{2}+\log2+s_0(\tanh\beta\la)+\beta\la\tanh\beta\la,
\lb{f_REMh}
\en
for $\beta < \betac$, while
\eq
-\beta f^{(2)}(\beta, \la)=
  \begin{cases}
   2 \beta\sqrt{\log 2}+\beta\la,&\la \ge 0;\\
  \beta \sqrt{\log2}+\beta\sqrt{s_0(\rho(\la))}+\beta\la\rho(\la),&\la < 0,
  \end{cases}
\lb{f_REMl}
\en
for $\beta \geq \betac$. 

We can now evaluate the left and right derivatives of $f^{(2)}(\beta,\la)$ at $\la=0$ by using the exact formula \rlb{f_REMh} and \rlb{f_REMl}.
The final results are
\eq
-\lim_{\la\dn0}\frac{\partial}{\partial\la}f^{(2)}(\beta,\la)=
\begin{cases}
0,&\beta<\betac;\\
1,&\beta\ge\betac,
\end{cases}
\lb{REMdef1}
\en
and
\eq
-\lim_{\la\up0}\frac{\partial}{\partial\la}f^{(2)}(\beta,\la)=0,
\lb{REMdef2}
\en
for any $\beta>0$.
They clearly indicate the two-replica free energy  $f^{(2)}(\beta,\la)$  is not differentiable with respect to $\la$ at $\la=0$ when $\beta\ge \betac$.
We also see that
\eq
\qjump=
\begin{cases}
0,&\beta<\betac;\\[3pt]
\dfrac{1}{2},&\beta\ge\betac.
\end{cases}
\lb{REMqjump}
\en
That the jump order parameter $q_{\rm jump}$ equals $\frac{1}{2}$ throughout the spin glass phase is consistent with the expression \rlb{REMPq} of the replica overlap distribution.
Since $P_\beta(q)$ has sharp peaks only at $q=0$ and $q=1$, the latter is amplified by the factor $e^{\beta N R(\bss^1,\bss^2)}$.
It might be unexpected that $q_{\rm jump}$ is still $\frac{1}{2}$ at the transition point $\beta=\betac$ (which is supported by a proof in the next section), especially if one notes that $P_{\betac}(q)=\delta(q)$ from \rlb{REMPq}.
In this case, we expect that the overlap distribution $P_{N,\betac}(q)$ for finite $N$ has a peak at $q=1$ that decays slowly with $N$.
Then the peak becomes visible because of the coupling factor $e^{\beta N R(\bss^1,\bss^2)}$.

Comparing \rlb{REMqbr} and \rlb{REMqjump}, we see that the conjectured inequality \rlb{qj>qbr}, $\qjump\ge\qbr$, is always valid in the REM.
The corresponding equality $\qjump=\qbr$ holds only for $\beta<\betac$ and at $\beta=2\betac$.

\subsection{Proof of the non-differentiability}
\label{s:REMp}
We supply the above exact calculation by a simple proof that the two-replica free energy is not differentiable at $\la=0$ in the spin glass phase.
More precisely, we prove
\eq
-\lim_{\la\dn0}\frac{\partial}{\partial\la}f^{(2)}(\beta,\la)=1,
\lb{REMrig1}
\en
and
\eq
-\lim_{\la\up0}\frac{\partial}{\partial\la}f^{(2)}(\beta,\la)\le1-\frac{\betac}{\beta},
\lb{REMrig2}
\en
for any $\beta\ge\betac$.
Although the bound \rlb{REMrig2} does not fully justify the expected behavior \rlb{REMdef2}, it is enough to establish that $f^{(2)}(\beta,\la)$ is not differentiable at $\la=0$.

As a starting point, note that \rlb{REMf2} implies
\eq
-\frac{1}{\la}\bigl\{f^{(2)}(\beta,\la)-f^{(2)}(\beta,0)\bigr\}=
\frac{1}{\la}\limN\frac{1}{\beta N}\bbE\log\frac{Z^{(2)}_N(\beta,\la)}{Z^{(2)}_N(\beta,0)}.
\lb{REMbasic}
\en
By only keeping the terms with $\bss^1=\bss^2$ in \rlb{REMf2}, we get the simple lower bound:
\eq
Z^{(2)}_N(\beta,\la)\ge\sum_{\bss\in\calC_N}e^{-\beta H_N^{(2)}(\bss,\bss;\la)}
=\sum_{\bss\in\calC_N}e^{-2\beta E_{\bss}+\beta\la\bss\cdot\bss}=e^{\beta\la N}\,Z_N(2\beta)
\lb{REMZ2>}
\en
Substituting this bound into \rlb{REMbasic} and assuming $\la>0$, we get
\eq
-\frac{1}{\la}\bigl\{f^{(2)}(\beta,\la)-f^{(2)}(\beta,0)\bigr\}\ge
1+\frac{1}{\la}\limN\frac{1}{\beta N}\bbE\log\frac{Z_N(2\beta)}{\{Z_N(\beta)\}^2},
\en
where we noted that $Z_N^{(2)}(\beta,0)=\{Z_N(\beta)\}^2$.
Now, from \rlb{RSBf1}, we see
\eq
\limN\frac{1}{\beta N}\bbE\log\frac{Z_N(2\beta)}{\{Z_N(\beta)\}^2}=2\{f(\beta)-f(2\beta)\}=0,
\en
where the final equality follows from \rlb{REMfsol}, and is valid for $\beta\ge\betac$.
By letting $\la\dn0$, we obtain
\eq
-\lim_{\la\dn0}\frac{\partial}{\partial\la}f^{(2)}(\beta,\la)\ge1.
\en
Since the left-hand side cannot exceed unity (because $|R|\le1$), we get the desired \rlb{REMrig2}.

From \rlb{REMH2}, \rlb{REMZ2}, and \rlb{REM<F>}, we see
\eq
\frac{Z^{(2)}_N(\beta,\la)}{Z^{(2)}_N(\beta,0)}
=\sbkt{e^{\beta\la N\,R}}^{(2)}_{N,\beta,0}
\ge e^{\beta\la N\,\sbkt{R}^{(2)}_{N,\beta,0}},
\en
where we used Jensen's inequality.
Substituting this bound into \rlb{REMbasic} and assuming $\la<0$, we get
\eq
-\frac{1}{\la}\bigl\{f^{(2)}(\beta,\la)-f^{(2)}(\beta,0)\bigr\}\le\limN\bbE\sbkt{R}^{(2)}_{N,\beta,0},
\en
which, with \rlb{REMR}, leads to the desired \rlb{REMrig2} if we let $\la\up0$.

\section{Griffiths-type theorem for a general first-order phase transition}
\label{s:gen1st}
By a straightforward extension of the proof of Theorem~\ref{t:Grgen}, one can prove a similar general theorem for the first-order phase transition in a non-random spin system.
Again the theorem is valid for any spin system with bounded classical spins and translation invariant interactions, but we shall discuss a basic class of the Ising model for simplicity.

Consider a general Hamiltonian $H_L^{(0)}(\bss)$ on $\LaL$ with translation invariant short-range interactions.
We can take, for example,
\eq
H_L^{(0)}(\bss)=-\sum_{x,y\in\LaL}K_{x,y}\sigma_x\sigma_y-\sum_{x,y,z\in\LaL}K'_{x,y,z}\sigma_x\sigma_y\sigma_z,
\en
where the interactions $K_{x,y}$ and $K'_{x,y,z}$ (with $x,y,z\in\bbZ^d$) are non-random and satisfy
\eq
K_{x,y}=K_{x+u,y+u},\quad K'_{x,y,z}=K'_{x+u,y+u,z+u},
\en
for any $u\in\bbZ^d$.
We also assume that $K_{x,y}=0$ for any $x,y$ such that $|x-y|>r_0$ and $K'_{x,y,z}=0$ for any $x,y,z$ such that $\max\{|x-y|,|y-z|,|z-x|\}>r_0$, where the range $r_0$ is a constant.
We define the thermal expectation value of any function $F(\bss)$ and the partition function by
\eq
\sbkt{F}_{L,\beta,h}\coloneqq\frac{1}{Z_L(\beta,h)}\sum_{\bss\in\calC_L}F(\bss)\,e^{-\beta \{H_L^{(0)}(\bss)-h\,\calO_L(\bss)\}},
\en
and
\eq
Z_L(\beta,h)\coloneqq\sum_{\bss\in\calC_L}e^{-\beta \{H_L^{(0)}(\bss)-h\,\calO_L(\bss)\}},
\en
respectively, where $\calO_L$ is the order parameter, i.e., the sum over $\LaL$ of the translated copies of an arbitrary local operator; an obvious example is the total magnetization
\eq
\calO_L(\bss)\coloneqq\sum_{x\in\LaL}\sigma_x.
\en
Then the specific free energy is defined by
\eq
f(\beta,h)\coloneqq-\limL\frac{1}{\beta L^d}\log Z_L(\beta,h).
\en

We define the order parameter that characterizes an extensive fluctuation of $\calO_L$ by
\eq
\mu_{\rm fluc}(\beta_0,h_0)\coloneqq\limsupL\frac{1}{L^d}\sqrt{\sbkt{(\calO_L)^2}_{L,\beta_0,h_0}-\bigl(\sbkt{\calO_L}_{L,\beta_0,h_0}\bigr)^2},
\en
and that characterizing the jump of the derivative of the free energy as
\eq
\mu_{\rm jump}(\beta_0,h_0)\coloneqq\frac{1}{2}\Bigl\{-\lim_{h\downarrow h_0}\frac{\partial}{\partial h}f(\beta_0,h)-\lim_{h\uparrow h_0}\frac{\partial}{\partial h}f(\beta_0,h)\Bigr\}.
\en
We then have the following.
\begin{theorem}[Griffiths-type theorem for a general first-order phase transition]\label{t:GrgenAp}
It holds for any $\beta_0$ and $h_0$ that
\eq
\mu_{\rm jump}(\beta_0,h_0)\ge\frac{1}{4}\bigl\{\mu_{\rm fluc}(\beta_0,h_0)\bigr\}^2.
\lb{mumuAp}
\en
\end{theorem}

The theorem establishes that whenever $\mu_{\rm fluc}(\beta_0,h_0)>0$ there exists a first order transition at $(\beta_0,h_0)$ such that the expectation value of the order parameter $o(\beta_0,h)\coloneqq\limL L^{-d}\sbkt{\calO_L}_{L,\beta_0,h}$ exhibits discontinuity as $h$ is varied.

The inequality \rlb{mumuAp} can be considerably improved when certain conditions are met.
If one has $\sbkt{\calO_L}_{L,\beta_0,h_0}=0$, then one can prove $\mu_{\rm jump}(\beta_0,h_0)\ge\mu_{\rm fluc}(\beta_0,h_0)/2$.
If one further has the symmetry $f(\beta_0,h_0+\delta)=f(\beta_0,h_0-\delta)$, then one gets  $\mu_{\rm jump}(\beta_0,h_0)\ge\mu_{\rm fluc}(\beta_0,h_0)$.

\paragraph*{Acknowledgments}
It is a pleasure to thank Aernout van Enter, Koji Hukushima, Harukuni Ikeda, Hidetoshi Nishimori, Yoshinori Sakamoto, Masafumi Udagawa, and Mizuki Yamaguchi for their valuable discussions.
The present research is partly supported by JSPS Grants-in-Aid for Scientific Research Nos.~21K03393 (C.I.) and 22K03474 (H.T.).

\medskip
\noindent
\paragraph*{Conflict of interest statement}
The authors declare no conflicts of interest.

\paragraph*{Data availability statement}
Data sharing not applicable to this article as no datasets were generated or analyzed
during the current study and article describes entirely theoretical research.

\end{document}